\documentclass[sigconf]{acmart}

\AtBeginDocument{%
  }

\setcopyright{acmcopyright}
\copyrightyear{2024}
\acmYear{2024}
\acmDOI{XXXXXXX.XXXXXXX}

\acmConference[AsiaCCS '24]{The 19th ACM ASIA Conference on Computer and Communications Security}{July 01--05,
  2024}{Singapore}
\acmPrice{15.00}
\acmISBN{978-1-4503-XXXX-X/18/06}

\usepackage{adjustbox}
\usepackage[linesnumbered,nofillcomment,ruled,vlined]{algorithm2e}
\usepackage{amsfonts}
\usepackage{amsmath}
\usepackage{amsthm}
\usepackage{booktabs}
\usepackage{colortbl}

\usepackage{enumitem}
\usepackage{epigraph}
\usepackage{graphicx}
\usepackage[htt]{hyphenat}
\usepackage[framemethod=TikZ]{mdframed}
\usepackage{microtype}
\usepackage{multirow}
\usepackage{natbib}
\usepackage{pifont}
\usepackage{siunitx}
\usepackage{soul}
\usepackage{subcaption}
\usepackage{tikz}
\usepackage{xcolor}
\usepackage{listings}

\PassOptionsToPackage{hyphens}{url}\usepackage{hyperref}

\usepackage[capitalize,sort&compress]{cleveref}

\theoremstyle{definition}

\DeclareSIUnit{\loc}{LoC}
\DeclareSIUnit{\year}{yr}
\DeclareSIUnit{\input}{input}

\newlist{inlineroman}{enumerate*}{1}
\setlist[inlineroman]{label=(\roman*)}

\newlist{inlinealph}{enumerate*}{1}
\setlist[inlinealph]{label=(\alph*)}

\newlist{researchq}{enumerate}{1}
\setlist[researchq]{
  label=\textbf{RQ~\arabic*},
  ref=\arabic*,
}
\crefname{researchqi}{RQ}{RQs}
\crefname{researchqi}{RQ}{RQs}

\definecolor{snsblue}{HTML}{4C72B0}
\definecolor{snsorange}{HTML}{DD8452}
\definecolor{snsgreen}{HTML}{55A868}
\definecolor{snsred}{HTML}{C44E52}
\definecolor{snspurple}{HTML}{8172B3}
\definecolor{snsbrown}{HTML}{937860}
\definecolor{snspink}{HTML}{DA8BC3}
\definecolor{snsgrey}{HTML}{8C8C8C}
\definecolor{snsgold}{HTML}{CCB974}
\definecolor{snscyan}{HTML}{64B5CD}

\colorlet{hlsnsblue}{snsblue!25}
\colorlet{hlsnsorange}{snsorange!25}
\colorlet{hlsnsgreen}{snsgreen!25}
\colorlet{hlsnsred}{snsred!25}
\colorlet{hlsnspurple}{snspurple!25}
\colorlet{hlsnsbrown}{snsbrown!25}
\colorlet{hlsnspink}{snspink!25}
\colorlet{hlsnsgrey}{snsgrey!25}
\colorlet{hlsnsgold}{snsgold!25}
\colorlet{hlsnscyan}{snscyan!25}

\definecolor{mplblue}{HTML}{1f77b4}
\definecolor{mplorange}{HTML}{ff7f0e}
\definecolor{mplgreen}{HTML}{2ca02c}
\definecolor{mplred}{HTML}{d62728}
\definecolor{mplpurple}{HTML}{9467bd}

\colorlet{hlmplblue}{mplblue!25}
\colorlet{hlmplorange}{mplorange!25}
\colorlet{hlmplgreen}{mplgreen!25}
\colorlet{hlmplred}{mplred!25}
\colorlet{hlmplpurple}{mplpurple!25}

\definecolor{set19c1}{HTML}{e41a1c} 
\definecolor{set19c2}{HTML}{377eb8} 
\definecolor{set19c3}{HTML}{4daf4a} 
\definecolor{set19c4}{HTML}{984ea3} 
\definecolor{set19c5}{HTML}{ff7f00} 
\definecolor{set19c6}{HTML}{ffff33} 
\definecolor{set19c7}{HTML}{a65628} 
\definecolor{set19c8}{HTML}{f781bf} 
\definecolor{set19c9}{HTML}{999999} 

\newrobustcmd{\myhl}[2]{{\sethlcolor{#1}\hl{\mbox{#2}}}}

\newcommand{\xmark}{\ding{55}\xspace}
\newcommand{\appx}{\ensuremath{\mathord{\sim}}}
\newcommand{\gt}{\ensuremath{\mathord{>}}}

\newcommand{\algoname}{\textsc{T-Scheduler}\xspace}
\newcommand{\algoworare}{\textsc{Rare}\textsuperscript{$-$}\xspace}
\newcommand{\algowrare}{\textsc{Rare}\textsuperscript{$+$}\xspace}
\newcommand{\algosample}{\textsc{Sample}\xspace}
\newcommand{\artifacturl}{\url{https://github.com/asiaccs2024-t-scheduler}\xspace}

\newcommand{\aflfast}{\textsc{AFLFast}\xspace}
\newcommand{\aflhier}{\textsc{AFL-Hier}\xspace}

\newcommand{\fuzzbench}{\textsc{FuzzBench}\xspace}
\newcommand{\entropic}{\textsc{Entropic}\xspace}
\newcommand{\kscheduler}{\texttt{K-Scheduler}\xspace}
\newcommand{\tortoisefuzz}{TortoiseFuzz\xspace}

\newcommand{\prog}{\ensuremath{\mathcal{P}}\xspace}
\newcommand{\covmapsize}{\ensuremath{K}\xspace}
\newcommand{\covmap}{\ensuremath{\mathcal{C}}\xspace}
\newcommand{\seedinput}{\ensuremath{\iota}\xspace}
\newcommand{\inputcovmap}{\ensuremath{\mathcal{C} (\prog, \seedinput) }\xspace}
\newcommand{\seedinputs}{\ensuremath{\mathcal{I}}\xspace}

\newcommand{\numtrials}{ten\xspace}
\newcommand{\fuzztime}{\qty{72}{\hour}\xspace}
\newcommand{\numfuzzbenchtargets}{19\xspace}

\newcommand{\numprograms}{35\xspace}
\newcommand{\numcomparedfuzzers}{four\xspace}
\newcommand{\numcomparedschedulers}{11\xspace}
\newcommand{\numaflschedulers}{eight\xspace}

\newcommand{\totalbugfuzztime}{\qty{16}{CPU\textrm{-}\year}\xspace}

\newcommand{\totalcovfuzztime}{\qty{19}{CPU\textrm{-}\year}\xspace}

\newcommand{\totalfuzztime}{\qty{35}{CPU\textrm{-}\year}\xspace}

\newcommand{\aflhieroverhead}{\ensuremath{2\times}\xspace}

\newcommand{\pvalue}{\ensuremath{p\textrm{-value}}\xspace}
\newcommand{\ci}[1]{{\scriptsize\color{darkgray}\ensuremath{\,\pm\,#1}}}
\newcommand{\greencell}{\cellcolor{hlsnsgreen}}
\newcommand{\whitecell}{\cellcolor{white}}
\newcommand{\shaderow}{\rowcolor{lightgray!10}}

\DontPrintSemicolon
\SetAlFnt{\footnotesize}
\LinesNumbered
\SetNlSty{}{}{}

\newenvironment{hlbox}[1]{
  \mdfsetup{
    linecolor=snsgrey,
    linewidth=2pt,
    roundcorner=2pt,
    frametitle={\colorbox{hlsnsgrey}{\space\textbf{#1}\space}},
    frametitleaboveskip=\dimexpr-\ht\strutbox\relax,
    innertopmargin=0pt,
    skipabove=0.5\baselineskip,
    skipbelow=0.5\baselineskip,
    leftmargin=5pt,
    rightmargin=5pt,
    nobreak=true,
  }
  \begin{mdframed}%
}{
  \end{mdframed}
}

\usetikzlibrary{chains}

\begin{document}

\title{Make out like a (Multi-Armed) Bandit: Improving the Odds of Fuzzer Seed Scheduling with \algoname}


\author{Simon Luo}
\affiliation{%
  \institution{The University of New South Wales}
  \country{Australia}
}
\email{simon.luo@unsw.edu.au}

\author{Adrian Herrera}
\affiliation{%
  \institution{Australian National University}
  \country{Australia}
}

\author{Paul Quirk}
\author{Michael Chase}
\affiliation{%
  \institution{Defence Science \& Technology Group}
  \country{Australia}
}

\author{Damith C.\ Ranasinghe}
\affiliation{%
  \institution{University of Adelaide}
  \country{Australia}
}

\author{Salil S.\ Kanhere}
\affiliation{%
  \institution{The University of New South Wales}
  \country{Australia}
}

\renewcommand{\shortauthors}{Luo et al.}

\begin{abstract}

Fuzzing is an industry-standard software testing technique that uncovers bugs in a target program by executing it with mutated inputs.
Over the lifecycle of a fuzzing campaign, the fuzzer accumulates inputs inducing new and interesting target behaviors, drawing from these inputs for further mutation and generation of new inputs.
This rapidly results in a large pool of inputs to select from, making it challenging to quickly determine the ``most promising'' input for mutation.
Reinforcement learning (RL) provides a natural solution to this \textit{seed scheduling} problem---\textit{a fuzzer can dynamically adapt its selection strategy by learning from past results}.
However, existing RL approaches are
\begin{inlinealph}
\item computationally expensive (reducing fuzzer throughput), and/or

\item require hyperparameter tuning (reducing generality across targets and input types).
\end{inlinealph}
To this end, we propose \algoname, a seed scheduler built upon multi-armed bandit theory to automatically adapt to the target.
Notably, our formulation does not require the user to select or tune hyperparameters and is therefore easily generalizable across different targets.
We evaluate \algoname over \totalfuzztime fuzzing effort, comparing it to~\numcomparedschedulers state-of-the-art schedulers.
Our results show that \algoname improves on these~\numcomparedschedulers schedulers on both bug-finding and coverage-expansion abilities.

\end{abstract}




\maketitle

\section{Introduction}
\label{sec:intro}

\epigraph{%
  ``\textit{Make out like a bandit}''.
  Idiom.
  To make a large profit.
}{%
  Merriam-Webster Dictionary
}

Fuzzing is a software testing technique for automatically finding bugs and vulnerabilities in a target program.
Fuzzers find bugs by mutating inputs to induce new behavior in the target.
Intuitively, mutated inputs are more likely to exercise corner cases in the target's behaviors, leading to bugs.
While most of these mutations do not lead to anything interesting, there remains a chance that the mutated input induces new and interesting target behaviors.

Intelligently selecting which inputs to mutate is critical for maximizing fuzzer effectiveness; inputs more likely to uncover new behaviors should be prioritized for mutation.
This prioritization of inputs is known as \emph{seed scheduling}\footnote{We use the term ``seed scheduling'', rather than ``seed selection'', to disambiguate it from the (offline) process of selecting an initial set of inputs to bootstrap the fuzzer.}~\cite{Manes:2019:FuzzingSurvey,wang2021reinforcement}.
Seed schedulers typically use heuristics to determine an input's position in the fuzzer's queue.
In a coverage-guided greybox fuzzer---the most common type of fuzzer---seed scheduling can be driven by a combination of:
\begin{inlineroman}
\item code coverage (inputs leading to new code uncover new behaviors);

\item input size (smaller inputs are faster to mutate);

\item execution time (inputs with shorter execution time mean more fuzzer iterations);

\item the number of times the input has been previously selected (avoiding local optima); and

\item similarity with other inputs (improving diversity).
\end{inlineroman}
However, seed scheduling is challenging because of a combination of the
\begin{inlinealph}
\item large number of inputs generated via mutation (and thus requiring prioritization),

\item large search space of the target, and

\item computational requirements (e.g., CPU time, memory, and storage).
\end{inlinealph}

Machine learning (ML)---in particular, reinforcement learning (RL)---is commonly applied to solve challenges in fuzzing~\cite{li2022deep, wang2020systematic, saavedra2019review, wang2021reinforcement, Wang:2021:SyzVegas, Bottinger:2018:DeepReinforcementFuzzing, Scott:2020:BanditFuzz, Godefroid:2017:LearnAndFuzz, Cheng:2019:SeedOptimization, Zong:2020:FuzzGuard, Chen:2020:Meuzz, Siddharth:2018:ThompsonFuzzing}.
Notably, RL has been used to adaptively learn seed scheduling strategies more likely to lead to increased code coverage.
In turn, this increases the likelihood of uncovering new bugs (after all, one cannot find bugs in code that is never executed).
However, integrating RL into fuzzing introduces two challenges: \emph{performance tradeoffs} and \emph{hyperparameter tuning}.

\paragraph*{Performance tradeoffs}
A fuzzer's iteration rate is the number of inputs the fuzzer executes per unit of time; the faster the fuzzer's iteration rate, the quicker the fuzzer can discover new and interesting behaviors.
However, balancing performance and ``cleverness'' in selecting the best input to mutate is difficult.
Moreover, RL algorithms require computational resources to train and evaluate, impacting a fuzzer's iteration rate.
Na\"{i}vely introducing RL into a fuzzer (notably, for seed scheduling) can increase run-time overhead without any performance improvement.
For example, we found \aflhier~\cite{wang2021reinforcement} (a fuzzer using RL for seed scheduling) introduced~$\gt\aflhieroverhead$ overhead over AFL++'s~\cite{Fioraldi:2020:AFLPlusPlus} heuristic-based scheduler without any improvement in fuzzing outcomes.

\paragraph*{Hyperparameter tuning}
RL algorithms use hyperparameters to configure their learning process.
The number of hyperparameters depends on the RL algorithm used.
For example, \aflfast~\cite{bohme2016coverage}, EcoFuzz~\cite{yue2020ecofuzz}, \aflhier~\cite{wang2021reinforcement}, and MobFuzz~\cite{Zhang:2022:MobFuzz} (fuzzers using RL in their seed schedulers) each have two hyperparameters.
Hyperparameters must be set before learning (and hence fuzzing) begins.
However, empirically selecting optimal hyperparameter values is time-consuming and difficult to generalize; optimal values are likely to vary across targets and input formats.
Suboptimal hyperparameter values reduce fuzzing performance.

\medskip
\noindent
We propose an RL approach that addresses these challenges and improves fuzzing outcomes.
Our approach models seed scheduling as a \emph{multi-armed bandit} (MAB) problem, which we solve using \emph{Thompson sampling}.
Thompson sampling allows us to adaptively and efficiently model the probability of the fuzzer uncovering new and interesting behaviors.
In doing so, the fuzzer can make more intelligent seed scheduling decisions with
\begin{inlinealph}
\item no hyperparameters to tune,

\item theoretical optimality guarantees~\cite{agrawal2017near}, and

\item constant-time overheads.
\end{inlinealph}
Our approach also uses a self-balancing mechanism to prioritize inputs covering rare paths and newly-discovered code.

We implement our RL-based seed scheduler, \algoname, in AFL++ (the current state-of-the-art coverage-guided greybox fuzzer) and evaluate it on \numprograms programs across two widely-used fuzzer benchmarks (Magma~\cite{Hazimeh:2020:Magma} and \fuzzbench~\cite{FuzzBench}).
Our evaluation shows that \algoname consistently improves on~\numcomparedschedulers state-of-the-art seed schedulers on~26 programs.
We contribute:

\begin{itemize}[leftmargin=*]
\item \textbf{A theoretical formulation of the seed scheduling problem.}
We formulate seed scheduling as a MAB problem.
Our formulation allows the fuzzer to prioritize inputs corresponding to newly-discovered target behaviors, based on learning the historical success of past seed inputs (\crefrange{sec:notation}{sec:input-selection}).

\item \textbf{An RL-based seed scheduler with no hyperparameters.} 
We design and implement \algoname, based on solving our MAB problem using Thompson sampling.
Using Thompson sampling means that we benefit from the inherent theoretical guarantee that model errors grow sublinearly---important for facilitating long fuzzing campaigns.
Moreover, our implementation has no hyperparameters, making it more generalizable to different targets and input formats.
We integrate our implementation into AFL++ (\cref{sec:input-selection}).

\item \textbf{An effective and generalizable seed scheduler.}
We evaluate \algoname by fuzzing real-world programs ($\gt\totalfuzztime$).
Our approach outperforms current state-of-the-art schedulers across both bug-finding and coverage-expansion metrics
(\cref{sec:evaluation}).

\item \textbf{Analysis of seed scheduler costs.}
Seed schedulers must carefully balance overhead costs and precision. 
To this end, we empirically analyze the cost of existing schedulers to understand their impact on fuzzing outcomes (\cref{sec:eval-overheads}).
\end{itemize}
We release our implementation and results at \artifacturl.

\section{Background}
\label{sec:background}

\subsection{Fuzzing}
\label{sec:fuzzing-background}

Fuzz testing (``fuzzing'') is a dynamic analysis for uncovering bugs in software.
In a security context, fuzzers have been wildly successful at discovering tens of thousands of security-critical vulnerabilities in widely-used code~\cite{Google:2023:OSSFuzz}.
Bugs are found by (rapidly) subjecting a target program to automatically-generated inputs.
The fuzzer generates inputs to expose and explore corner cases in the target not considered by the developer.
Intuitively, it is in these corner cases where bugs are most likely to lie.

\begin{figure}
\centering
\small
\includegraphics[width=\linewidth,keepaspectratio]{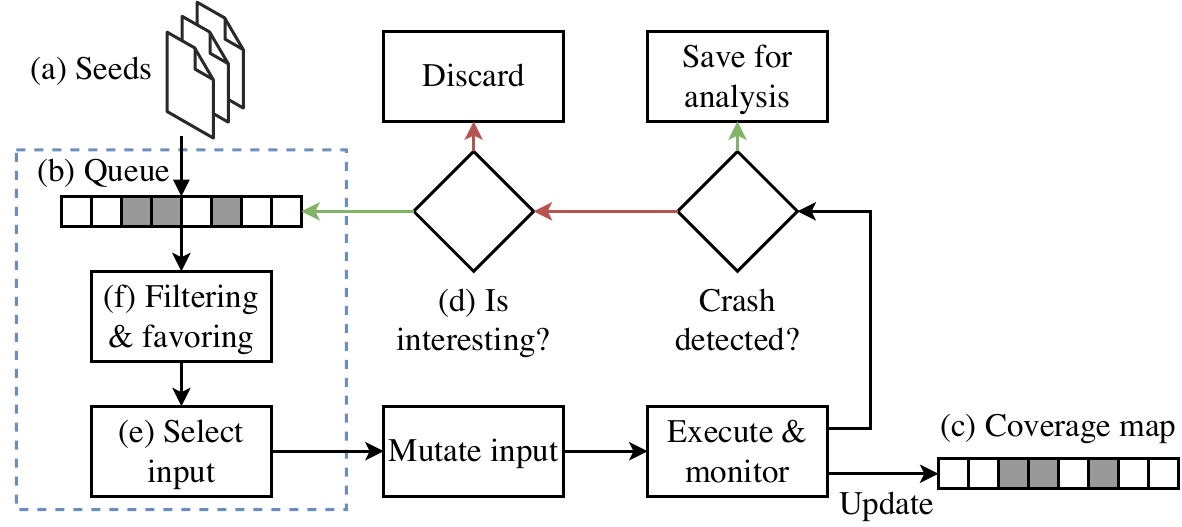}
\caption{Overview of Greybox Fuzzing.}
\label{fig:overview_greybox}
\end{figure}

\Cref{fig:overview_greybox} illustrates the fuzzing process.
A fuzzing campaign begins with curating a corpus of ``well-formed'' inputs.
These inputs are commonly exemplar input data accepted by the target (\cref{fig:overview_greybox}(a))~\cite{Rebert:2014:SeedSelection,Herrera:2021:SeedSelection}.
At run time, the fuzzer maintains a queue---initially populated from this curated corpus---from which an input is selected to fuzz (\cref{fig:overview_greybox}(b)).
The selected input is mutated and fed into the target to expose corner cases in program behavior.
How does the fuzzer select these inputs?

\emph{Greybox} fuzzers use lightweight instrumentation to track code executed (or ``covered'') by the target (in contrast, \emph{blackbox} fuzzers have no internal view of the target).
The fuzzer records code coverage in a \emph{coverage map}, which tracks the number of times a particular coverage element---typically, an edge in the target's control-flow graph (CFG)---is executed (\cref{fig:overview_greybox}(c)).
By tracking code coverage, the fuzzer can determine if the mutated input triggers new target behaviors.
Inputs triggering new behaviors are saved back into the queue; otherwise, the input is discarded (\cref{fig:overview_greybox}(d)).
This guides the fuzzer to prioritize inputs leading to new program behaviors (as captured by the instrumentation).
Selecting and prioritizing inputs for mutation is handled by the \emph{seed scheduler} (\cref{fig:overview_greybox}(e)).

\subsection{Seed Scheduling}
\label{sec:seed-scheduling-background}

An efficient seed scheduler (the blue box in \cref{fig:overview_greybox}) must satisfy two competing constraints.
First, the seed scheduler must \emph{select the most-promising input for mutation}; i.e., the input most likely to cover new program behaviors.
Second, it must make this selection from a (potentially) large queue with \emph{minimal run-time overhead}.
AFL-based~\cite{Zalewski:2015:AFL} fuzzers solve this problem by bounding the input queue to a fixed size via an \emph{input filtering and favoring} phase~\cite{Wang:2020:TortoiseFuzz}.

\paragraph*{Input filtering and favoring}
The fuzzer only retains inputs that induce new and interesting behaviors in the target; e.g., an input that covers a new element in the coverage map.
These inputs are made available for future mutations, potentially uncovering more new behaviors.
AFL ``favors'' an input if it is the fastest and smallest input for any of the coverage map elements~\cite{bohme2016coverage} (\cref{fig:overview_greybox}(f)).
Thus, tracking favored inputs gives the fuzzer a minimal set of inputs (that are both small and fast) covering all of the elements seen in the coverage map so far (approximating a \emph{weighted minimum set cover}, with size and speed as weights~\cite{Fioraldi:2020:AFLPlusPlus}).
Maintaining a set of favored inputs implicitly reduces the seed scheduling problem from an unbounded number of inputs (the union of the initial seed corpus and the inputs generated so far) to a bounded number of inputs: the number of favored inputs; i.e., the coverage map's size.

\medskip
\noindent
After filtering and favoring, the scheduler has a bounded number of inputs from which to select an input to fuzz.
From this, the scheduler selects the ``best'' input to mutate.
For example, AFL selects an input based on a score calculated using a set of heuristics.
These heuristics calculate a performance score based on an input's:
\begin{inlineroman}
\item coverage;

\item execution time (faster inputs are preferred);

\item ``depth'' (i.e., the number of inputs mutated to reach the given input); and

\item the fuzzer's run time (newer inputs are prioritized).
\end{inlineroman}
A \emph{power schedule}~\cite{bohme2016coverage,Fioraldi:2020:AFLPlusPlus} then distributes fuzzing time across inputs by scaling the performance score based on the number of times the input has been selected, biasing fuzzing time to less-fuzzed inputs.
Higher energy means the fuzzer spends more time mutating the corresponding (favored) input.

Notably, AFL's heuristics are based on intuition and experimentation.
Other fuzzers (e.g., EcoFuzz~\cite{yue2020ecofuzz}, \aflhier~\cite{wang2021reinforcement}, MobFuzz~\cite{Zhang:2022:MobFuzz}, and \kscheduler~\cite{She:2022:KScheduler}) also rely on hyperparameters that must be tuned per target to achieve optimal results.
In contrast, our approach (described in \cref{sec:approach}) replaces these heuristics with an~RL algorithm that uses run-time statistics to dynamically learn and adapt a seed schedule.
Moreover, our approach has no hyperparameters to tune, leading to more efficient and informed input selection.

\subsection{Reinforcement Learning}
\label{sec:rl-background}

Reinforcement learning (RL) is an ML paradigm that trains an agent by observing changes in state and rewarding the selected actions~\cite{sutton2018reinforcement}.
The agent aims to select the best action to maximize the cumulative reward.
However, the expected reward for each action is often unknown and must be learned dynamically via experimentation.
This experimentation leads to a trade-off between \emph{exploiting} what is already known and \emph{exploring} territory.

An RL algorithm is typically defined in terms of \emph{states}, \emph{actions}, and \emph{rewards}.
The state is a set of variables describing the environment.
Based on the current state, the agent
\begin{inlinealph}
\item selects an action to perform, and

\item receives feedback on its selection in the form of a reward.
\end{inlinealph}
The agent's objective is thus to maximize the cumulative reward over a given time.

These concepts apply naturally to fuzzing.
In particular, the fuzzer's seed scheduler must select an input (to mutate) from a pool of constantly-changing possibilities.
Ideally, this selection maximizes the discovery of new code, or (ideally) a bug.
\emph{The seed scheduler must balance exploring the input queue and exploiting the input uncovering the most code}.
This requires a careful trade-off between making intelligent input prioritization decisions and maintaining the fuzzer's iteration rate.
We satisfy this trade-off by formulating seed scheduling as a \emph{multi-armed bandit}.

\subsubsection{Multi-Armed Bandit}
\label{sec:mab}

The multi-armed bandit (MAB) is a well-explored RL problem focusing on the trade-off between exploration and exploitation~\cite{lattimore2020bandit}.
Given a state, the agent selects an action~$a_{k} \in \mathcal{A}$ at each time step~$t \in [1, \ldots, T]$.
The agent's goal is to maximize the cumulative reward (by performing a sequence of actions) over~$T$.

The classic MAB involves~$K$ slot machines (``bandits''), where each~$k \in [1, \ldots, K]$ has an unknown probability~$\theta_k$ of paying out when played.
At each time step~$t$, the player (i.e., agent) selects a slot machine to play.
Once played, the player is either rewarded with a payout (with probability~$\theta_k$) or receives nothing (with probability~$1 - \theta_k$).
Naturally, any rational player would focus on the bandit paying out the most (thus achieving their goal of maximizing the cumulative reward).
Unfortunately, \emph{this information is unknown to the player}.
Consequently, the player must trade-off between \emph{exploiting} the bandit with the (current) highest expected payout and \emph{exploring} different bandits to learn more about the probability~$\theta_{k}$ (in the hope of finding a higher payout).
What is the best strategy for selecting between exploration and exploitation?

\emph{Thompson sampling}~\cite{thompson1933likelihood} is a popular approach for addressing this exploration/exploitation trade-off.
This popularity is due to simplicity, fast execution time, and optimality guarantees (ensuring errors grow sublinearly over time~\cite{agrawal2017near}).
Consequently, we adopt Thompson sampling in \algoname, our RL-based seed scheduler.

\section{Approach}
\label{sec:approach}

Our seed scheduler, \algoname, formulates greybox fuzzing as a Beta-Bernoulli bandit, which we solve with Thompson sampling.
We first provide a high-level description of the \algoname algorithm (\cref{sec:algorithm}) and a motivating example (\cref{sec:motivating-example}), followed by a mathematical formulation of our Beta-Bernoulli bandit model (\crefrange{sec:mab-scheduler}{sec:input-selection}).

\subsection{Notation and Definitions}
\label{sec:notation}

A fuzzer measures its progress fuzzing target~\prog in a coverage map~$\inputcovmap \in \mathbb{N}^\covmapsize$ of size~$K$.
Typically, each \emph{feature}~$x \in \inputcovmap$ records the number of times a particular edge in the target's control-flow graph (CFG)\footnote{
  Some fuzzers eschew edge coverage for other coverage metrics (e.g., context-sensitive edge or data flow).
  Our approach is agnostic to the underlying coverage metric.}
is executed by an input $\seedinput \in \seedinputs$, where~\seedinputs is the set of inputs representing the union of the initial corpus and the set of inputs generated by the fuzzer.
We refer to this count as a \emph{hit count} (given by the function \texttt{hit\_count}).

\emph{Feature rareness} prioritizes features covered less by \seedinputs.
We adopt the definition by \citet{wang2021reinforcement}, where feature rareness is the inverse of~$x$'s hit count.
Feature hit counts and rareness are defined for each feature in the coverage map and the inputs generated so far.
These definitions enable a seed scheduler that prioritizes both newly-discovered and hard-to-reach code.

\subsection{The \algoname Algorithm}
\label{sec:algorithm}

\begin{algorithm}
\caption{\algoname.}
\label{alg:RLFuzzing}

\SetKwFunction{FUpdatePosterior}{UpdatePosterior}
\SetKwFunction{FSelectInput}{SelectInput}
\SetKwFunction{FIsInteresting}{IsInteresting}
\SetKwFunction{FEnumerate}{Enumerate}
\SetKwFunction{FFavoredInputs}{FavoredInputs}

\SetKwProg{Fn}{Function}{}{}

$\boldsymbol{\alpha} \leftarrow \{ 1\, |\, k \in [1, \ldots, K] \}$\;\label{lin:alpha-init}
$\boldsymbol{\beta} \leftarrow \{ \,1 |\, k \in [1, \ldots, K] \}$\;\label{lin:beta-init}
\BlankLine

\Fn{\FUpdatePosterior{\inputcovmap,~$\boldsymbol{\alpha}, \boldsymbol{\beta}$}} {\label{lin:func_update_posterior_start}
\For{$k, x \in \inputcovmap$} {
\If{$x \ne 0$} {
\eIf{$\FIsInteresting(\seedinput)$} {
$\alpha_{k} \gets \alpha_{k} + 1$\;\label{lin:alpha}
} {
$\beta_{k} \gets \beta_{k} + 1$\;\label{lin:beta}
}
}
}
\KwRet {$\boldsymbol{\alpha}, \boldsymbol{\beta}$}
}\label{lin:func_update_posterior_end}

\BlankLine

\Fn{\FSelectInput{$\boldsymbol{\alpha}, \boldsymbol{\beta}$}}{\label{lin:func_select_next_input_start}
\For{$k \gets [1 , \ldots , K]$} {
$\hat{\theta}_{k} \sim \mathrm{Beta} (\alpha_{k}, \beta_{k})$\;\label{lin:beta_distribution}
$\psi_{k} \sim \mathrm{Beta} (\alpha_{k} + \beta_{k} , \alpha_{k}^2)$\;\label{lin:rareness_correction}
}
$a_{t} \gets \arg \max [\psi_{1} \hat{\theta}_{1} , \ldots, \psi_{K} \hat{\theta}_{K}]$\;\label{lin:select_input}
$\seedinputs^{(t+1)} \gets \FFavoredInputs(a_{t})$\;\label{lin:top_rated}

\KwRet {$\seedinputs^{(t+1)}$}
}\label{lin:func_select_next_input_end}
\end{algorithm}

We present the \algoname algorithm in \cref{alg:RLFuzzing}.
It consists of two functions: \texttt{UpdatePosterior} and \texttt{SelectInput}.

The \texttt{UpdatePosterior} function---called each time an input~\seedinput is executed---uses two~$K$-length vectors---$\boldsymbol{\alpha}$ and~$\boldsymbol{\beta}$ (\cref{lin:alpha,lin:beta})---to store the number of times a coverage map feature is hit or missed, respectively.
Each element~$\alpha_{k}$ and~$\beta_{k}$ (where~$k \in [1, \ldots, K]$) represents the number of times~\seedinput hits or misses~$x \in \inputcovmap$.
These vectors are used to compute a probability distribution for each coverage map feature, modeling the probability of an input inducing new behaviors in~\prog.

The \texttt{SelectInput} function is called when the queue has been exhausted.
It samples $\boldsymbol{\theta} = [\theta_{1}, \ldots , \theta_{K}]$ from~$K$ Beta distributions (\cref{lin:beta_distribution}) to determine the ``best'' input to fuzz (see \cref{sec:mab-scheduler}).
However, using only~$\boldsymbol{\theta}$ leads the fuzzer to repeatedly select the same inputs, because it implicitly penalizes rarely-covered features in the coverage map.
Thus, we introduce a \emph{correction factor}~$\boldsymbol{\psi} = [\psi_{1}, \ldots, \psi_{K}]$ (\cref{lin:rareness_correction}) based on feature rareness to penalize frequently-covered coverage map features (see \cref{sec:rareness-correction}).

Finally, the next input to fuzz is chosen from \texttt{FavoredInputs} using~$\hat{\theta}_{k}$ and~$\psi_{k}$ (\crefrange{lin:select_input}{lin:top_rated}; see \cref{sec:input-selection}).
\texttt{FavoredInputs} stores a single ``best'' input corresponding to each feature in the coverage map.
Per \cref{sec:seed-scheduling-background}, the fuzzer determines this input by a combination of execution time, input size, and the number of times the input has been fuzzed.

\subsubsection{Motivating Example}
\label{sec:motivating-example}

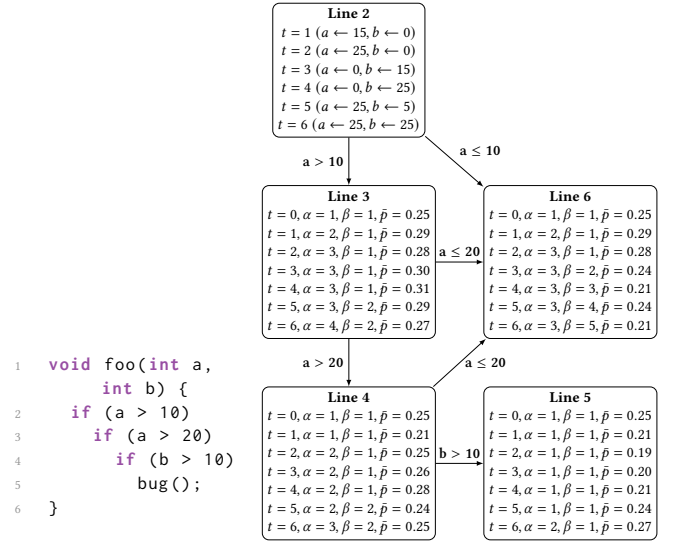
\begin{figure}
\begin{subfigure}[B]{0.34\linewidth}
\begin{lstlisting}[language=C]
void foo(int a, int b) {
  if (a > 10)
    if (a > 20)
      if (b > 10)
        bug();
}
\end{lstlisting}
\label{fig:example-code}
\caption{The program.
The parameters~\texttt{a} and~\texttt{b} are derived from user input.}
\end{subfigure}%
\hfill
\begin{subfigure}[B]{0.62\linewidth}
\centering
\begin{adjustbox}{width=\linewidth}
\begin{tikzpicture}[auto,
                      node distance = 10mm,
                      start chain = going below,
                      box/.style = {draw,rounded corners,fill=white,
                            on chain,align=center}]

    \node[box] (init)    {\textbf{Line 2} \\
                          $t=1$ $(a \leftarrow 15, b \leftarrow 0)$ \\
                          $t=2$ $(a \leftarrow 25, b \leftarrow 0)$ \\
                          $t=3$ $(a \leftarrow 0, b \leftarrow 15)$ \\
                          $t=4$ $(a \leftarrow 0, b \leftarrow 25)$ \\
                          $t=5$ $(a \leftarrow 25, b \leftarrow 5)$ \\
                          $t=6$ $(a \leftarrow 25, b \leftarrow 25)$};
    \node[box, below=of init] (a1)    {\textbf{Line 3} \\
                                              $t=0,\alpha=1,\beta=1,\bar{p}=0.25$ \\
                                              $t=1,\alpha=2,\beta=1,\bar{p}=0.29$ \\
                                              $t=2,\alpha=3,\beta=1,\bar{p}=0.28$ \\
                                              $t=3,\alpha=3,\beta=1,\bar{p}=0.30$ \\
                                              $t=4,\alpha=3,\beta=1,\bar{p}=0.31$ \\
                                              $t=5,\alpha=3,\beta=2,\bar{p}=0.29$ \\
                                              $t=6,\alpha=4,\beta=2,\bar{p}=0.27$};
    
    \node[box, below=of a1] (a2)    {\textbf{Line 4} \\
                                     $t=0,\alpha=1,\beta=1,\bar{p}=0.25$ \\
                                     $t=1,\alpha=1,\beta=1,\bar{p}=0.21$ \\
                                     $t=2,\alpha=2,\beta=1,\bar{p}=0.25$ \\
                                     $t=3,\alpha=2,\beta=1,\bar{p}=0.26$ \\
                                     $t=4,\alpha=2,\beta=1,\bar{p}=0.28$ \\
                                     $t=5,\alpha=2,\beta=2,\bar{p}=0.24$ \\
                                     $t=6,\alpha=3,\beta=2,\bar{p}=0.25$};
    
    \node[box, right=of a2] (b)    {\textbf{Line 5} \\
                                    $t=0,\alpha=1,\beta=1,\bar{p}=0.25$ \\
                                    $t=1,\alpha=1,\beta=1,\bar{p}=0.21$ \\
                                    $t=2,\alpha=1,\beta=1,\bar{p}=0.19$ \\
                                    $t=3,\alpha=1,\beta=1,\bar{p}=0.20$ \\
                                    $t=4,\alpha=1,\beta=1,\bar{p}=0.21$ \\
                                    $t=5,\alpha=1,\beta=1,\bar{p}=0.24$ \\
                                    $t=6,\alpha=2,\beta=1,\bar{p}=0.27$};
    
    \node[box, right=of a1] (join)    {\textbf{Line 6} \\
                                            $t=0,\alpha=1,\beta=1,\bar{p}=0.25$ \\
                                            $t=1,\alpha=2,\beta=1, \bar{p}=0.29$ \\
                                            $t=2,\alpha=3,\beta=1,\bar{p}=0.28$ \\
                                            $t=3,\alpha=3,\beta=2,\bar{p}=0.24$ \\
                                            $t=4,\alpha=3,\beta=3,\bar{p}=0.21$ \\
                                            $t=5,\alpha=3,\beta=4,\bar{p}=0.24$ \\
                                            $t=6,\alpha=3,\beta=5,\bar{p}=0.21$};

    \begin{scope}[rounded corners,-latex]
        \path
        (init) edge[] (join) (init) -- (join) node[midway,above right] (a <= 10) {$\mathbf{a \le 10}$}
        (init) edge[] (a1) (init) -- (a1) node[midway, left] (a > 10) {$\mathbf{a > 10}$}

        (a1) edge[] (a2) (a1) -- (a2) node[midway, left] (a > 20) {$\mathbf{a > 20}$}
        
        (a1) edge (join) (a1) -- (join) node[midway, above]  (a < 20) {$\mathbf{a \le 20}$}
        (a2) edge[] (join) (a2) -- (join) node[midway, right] (a < 20) {$\mathbf{a \le 20}$} 

        (a2) edge[] (b) (a2) -- (b) node[midway] (b > 10) {$\mathbf{b > 10}$}
        ;
    \end{scope}
\end{tikzpicture}
\end{adjustbox}
\caption{CFG showing the changing parameters~$\alpha_{k}$ and~$\beta_{k}$ at each time step~$t$.
The mean probability for each time step is shown as~$\bar{p}$.}
\end{subfigure}
\caption{
An example showing how \algoname updates its parameters after each input is executed.
Six different values for~\texttt{a} and~\texttt{b} (corresponding to six different inputs) are shown in the root node.
The parameter~$\alpha$ is incremented when the input discovers new program behavior, otherwise~$\beta$ is incremented.
The probability~$\bar{p}_k$ is the normalized mean of the Beta distribution in \cref{lin:beta_distribution} in \cref{alg:RLFuzzing} using \cref{eqn:scheduling_probability}.
}
\label{fig:toy_example_algo}
\end{figure}

We use the example in \cref{fig:toy_example_algo} to illustrate \algoname's approach.
\algoname is an \emph{adaptive} scheduler, meaning the probability of selecting a given input (for mutation) changes as the scheduler receives coverage feedback from the fuzzer.
In this example, coverage feedback is used to update~$\alpha$ (\cref{lin:alpha} in \cref{alg:RLFuzzing}) and~$\beta$ (\cref{lin:beta}), storing the positive and negative feedback of past scheduling decisions, respectively.
The scheduler uses the updated parameters~$\alpha$ and~$\beta$ to assign a probability for selecting an input (\cref{lin:beta_distribution}).
We show the mean of this probability distribution
\begin{align}
\label{eqn:scheduling_probability}
\bar{p}_{k} = \frac{\mathbb{E} [\mathrm{Beta} (\alpha_{k}, \beta_{k})] }{\sum_{k=1}^{K} \mathbb{E} [\mathrm{Beta} (\alpha_{k}, \beta_{k})]} ,
\end{align}
because of the difficulty visualizing~$\hat{\theta}_{k}$ in \cref{fig:toy_example_algo}.
Importantly, \cref{alg:RLFuzzing} does not need to compute~$\bar{p}_{k}$ and is only used to visualize how the mean of the probability distribution changes over time.

\Cref{fig:toy_example_algo} uses six example inputs to show how the parameters are updated.
Inputs discovering new program behaviors are stored in the corresponding node and~$\bar{p}$ is the probability of the input being selected by the scheduler.
This assumes that prior inputs that discovered new program behaviors are more likely to discover new program behaviors.
At~$t=1$, the input covers lines~3 and~6 (\cref{fig:example-code}) for the first time, so~$\alpha$ is incremented and~$\bar{p}$ increases.
At~$t=2$, the input covers lines~3,~4, and~6.
Line~4 is covered for the first time and~$\alpha$ is incremented for lines~3,~4, and~6.
Notably,~$\bar{p}$ in line~4 increases but~$\bar{p}$ decreases at lines~3 and~6 because we favor rarer paths.
The input covers line~6 at~$t=3$ and~$t=4$.
Line 6 has already been covered, so~$\beta$ is incremented and~$\bar{p}$ decreases (because we penalize inputs that do not uncover new program behaviors).
Similarly, at~$t=5$ the input covers lines~3,~4, and~6, which have also been covered by previous inputs.
Thus,~$\bar{p}$ decreases at these lines while~$\bar{p}$ increases for lines~2 and~6.
Finally, line~5 is covered for the first time at~$t=6$.
This results in~$\alpha$ increasing at lines~3,~4, and~5 and~$\bar{p}$ changing disproportionally to favor rarer paths.

\begin{figure}
\centering
\begin{subfigure}[t]{0.33\linewidth}
  \centering
  \includegraphics[width=\linewidth]{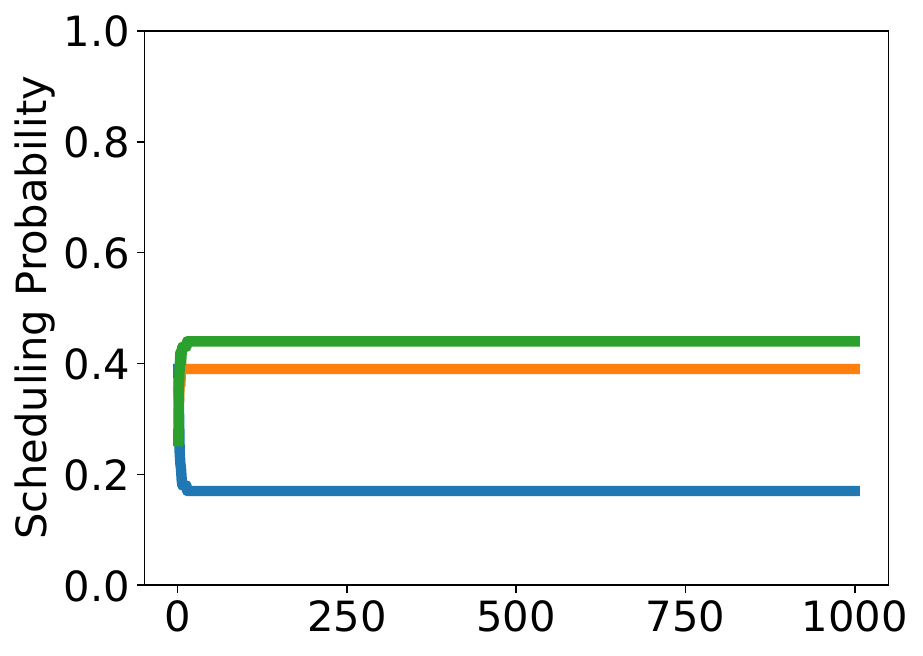}
  \caption{Greedy.}
  \label{fig:toy_greedy}
\end{subfigure}%
\hfill
\begin{subfigure}[t]{0.33\linewidth}
  \centering
  \includegraphics[width=\linewidth]{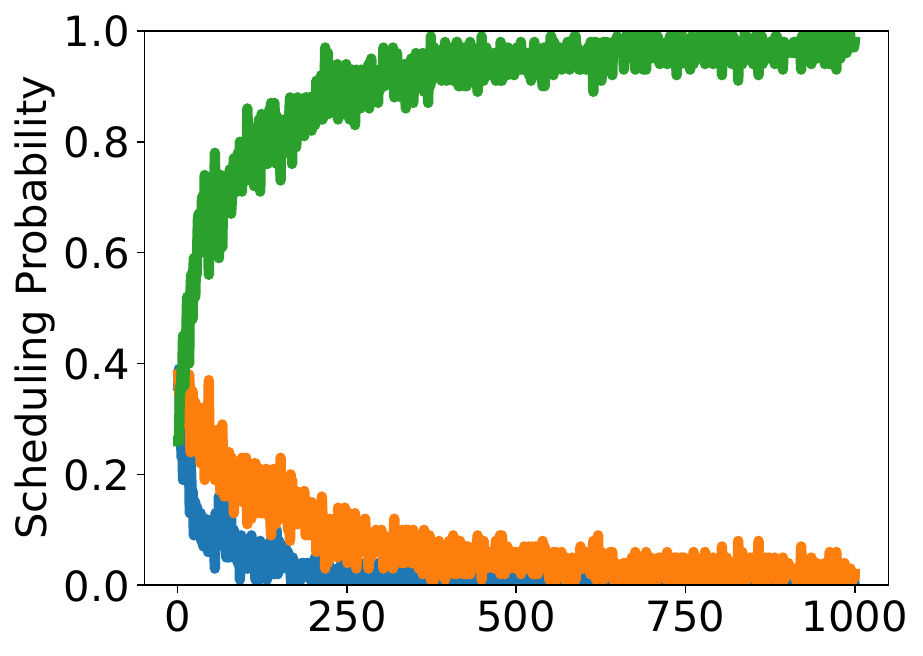}
  \caption{\algoname.}
  \label{fig:toy_thompson}
\end{subfigure}%
\hfill
\begin{subfigure}[t]{0.33\linewidth}
  \centering
  \includegraphics[width=\linewidth]{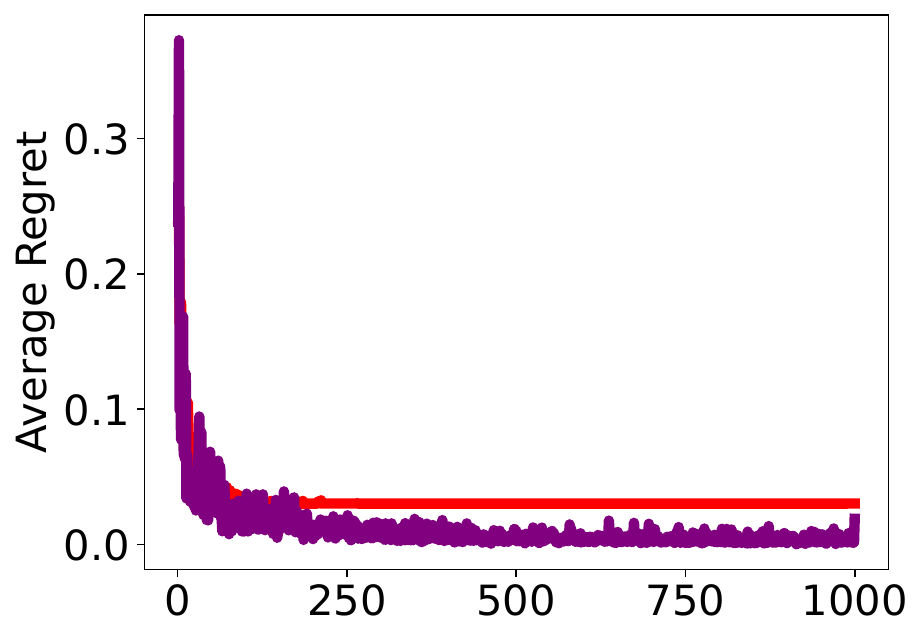}
  \caption{Regret.}
  \label{fig:toy_regret}
\end{subfigure}
\caption{An example scheduler demonstrating the probability of an input being selected over~\num{1000} iterations (the $x$-axis is iteration count).
The regret quantifies the error by taking the difference between the optimal decision and the input selected.
The inputs~\myhl{hlmplblue}{$\seedinput_{1}$},~\myhl{hlmplorange}{$\seedinput_{2}$}, and~\myhl{hlmplgreen}{$\seedinput{3}$} have latent probabilities of~0.7,~0.8, and~0.9, respectively.
The \myhl{hlmplred}{greedy} algorithm has a constant regret, while the \myhl{hlmplpurple}{\algoname} algorithm's regret approaches zero over time.
Highlight colors reflect those in the plots.}
\label{fig:toy_greed_vs_rl_example}
\end{figure}

The scheduler selects an input to executed based on the probability that mutating the input will lead to new program behaviors.
This probability is unknown.
Thus, inputs with a higher probability for discovering new program behaviors must be estimated and prioritized by the scheduler.
\Cref{fig:toy_greed_vs_rl_example} illustrates the impact different scheduling algorithms have on these estimates.
Here, we assume that a scheduler has three inputs to select from---$\seedinput_{1}$,~$\seedinput_{2}$, and~$\seedinput_{3}$---with probabilities~0.7,~0.8, and~0.9 that a mutation will discover new program behavior (which is unknown and needs to be estimated by the scheduler in a fuzzing campaign), respectively to demonstrate the behavior of the algorithm.\footnote{In a fuzzing campaign these probabilities change as new program behaviors are discovered.}
Here, we can see input $\seedinput_{3}$ should be selected because it has the highest probability to discover new program behavior.
But the scheduler does not know these values and is required to estimate them to assign a scheduling probability to each input.
\Cref{fig:toy_greedy,fig:toy_thompson} show the convergence of $\bar{p}$ for two scheduling approaches---a greedy algorithm and \algoname---when selecting inputs using the scheduling probabilities ($\bar{p}$ or $\hat{\theta}$).
\algoname allows for sub-optimal decisions at a given time by sampling from a distribution (\cref{lin:beta_distribution} in \cref{alg:RLFuzzing}) meaning the input with the highest $\bar{p}$ is not always selected.
These sub-optimal decisions allow the model to ``explore'' early on in the fuzzing campaign and ``exploit'' inputs when it has more information.
The greedy approach shown in \cref{fig:toy_greedy} always tries to select the highest scheduling probability which does not lead to the optimal solution.
\Cref{fig:toy_regret} illustrates the regret of \algoname converging to zero over time leading to the optimal solution while a constant regret remains for the greedy approach.
In this example, this can only be achieved if the scheduler always picks $\seedinput_{3}$.
Current state-of-the-art schedulers (e.g., \aflfast~\cite{bohme2016coverage}, \entropic~\cite{bohme2020boosting}, and \tortoisefuzz~\cite{Wang:2020:TortoiseFuzz}) use a greedy approach to make decisions.
\algoname makes more optimal decisions (compared to greedy algorithms), particularly in long fuzzing campaigns.

With the \algoname algorithm presented, we now turn our attention to the probabilistic modeling that underpins our approach.

\subsection{Adapting the Beta-Bernoulli Bandit}
\label{sec:mab-scheduler}

Thompson sampling frames exploration and exploitation as a Bayesian posterior estimation, choosing an action that maximizes a reward based on a randomly-drawn prior belief (\cref{sec:mab}).
We assume that information derived from previous inputs can be used to improve input scheduling in the future.
The reward function provides feedback to the fuzzer after each input~\seedinput is executed.
In particular, the fuzzer is (positively) rewarded for discovering new target behaviors, and penalized otherwise.

We begin our formulation with a~$K$-armed bandit; i.e., there are~$K$ actions for the fuzzer to choose from.
Here,~$K$ is the size of \inputcovmap, and each entry of the coverage map corresponds to a favored input in a many-to-one relationship (\cref{sec:seed-scheduling-background}).
We define a fuzzing campaign with respect to a time step~$t \in [1, \ldots, T]$, where~$T$ is the length of the campaign (i.e., the number of executed inputs).
Each input~$\seedinput^{(t)}$ at time step~$t$ will produce a coverage map~$\inputcovmap^{(t)}$.
After performing action~$k \in [ 1 , \ldots K ]$ the fuzzer is rewarded by:
\begin{equation}
\label{eqn:reward_function}
r_{k}^{(t)} := 
\begin{cases}
1, & \text{if } x_{k}^{(t)} \ne 0 \text{ and } \seedinput^{(t)} \text{ is interesting} , \\
0, & \text{if } x_{k}^{(t)} \ne 0 ,
\end{cases}
\end{equation}
where~$x_{k}^{(t)} \in \inputcovmap^{(t)}$ is the coverage map feature at index~$k$.
For each time step~$t$ the reward is represented as the vector~$\mathbf{r}^{(t)} = [r_{1}^{(t)}, \ldots , r_{K}^{(t)}]$.
The fuzzer is rewarded for inducing interesting behaviors in the target (e.g., uncovering new code). 
Intuitively, this ensures the scheduler selects inputs that are more likely to induce new behaviors in~\prog.

\subsubsection{Estimating Probabilities and Rewarding the Fuzzer}

The probability that the fuzzer generates an input (by mutating the current input) inducing new behaviors is~$\boldsymbol{\theta} = [ \theta_{1} , \ldots , \theta_{K}]$.
Importantly,~$\boldsymbol{\theta}$ is unknown and must be estimated over time through experimentation.
Per \cref{sec:rl-background}, learning~$\boldsymbol{\theta}$ requires a careful balance between exploration and exploitation to maximize the cumulative reward over~$T$.
The estimated~$\boldsymbol{\theta}$ guides the seed scheduler to select the next (best) input to fuzz.

We design the reward function in \cref{eqn:reward_function} such that~$\theta_{k}$ (the \emph{posterior distribution}) can be estimated using the Beta-Bernoulli bandit.
In this setting,~$\theta_{k}$ is the probability that the input induces new behaviors in~\prog, and is estimated by a Bernoulli distribution with observations~$r^{(t)}_{k}$ (the \emph{likelihood}) and a Beta distribution over~$\hat{\theta}^{(t)}_{k}$ (the \emph{prior distribution}); i.e.,
\begin{equation}
\label{eqn:bernoulli_likelihood}
r_{k}^{(t)} \sim p(r_{k}^{(t)} | \hat{\theta}^{(t)}_{k}) = \mathrm{Bernoulli} ( \hat{\theta}^{(t)}_{k} )
\end{equation}
and
\begin{equation}
\label{eqn:beta_prior}
\hat{\theta}^{(t)}_{k} \sim p(\hat{\theta}^{(t)}_{k} ; \alpha_{k} , \beta^{(t)}_{k}) = \mathrm{Beta} ( \alpha_{k}^{(t)} , \beta_{k}^{(t)} ).
\end{equation}

The prior distribution is modeled as a Beta distribution with parameters~$\boldsymbol{\alpha}^{(t)} = [ \alpha^{(t)}_{1} , \ldots , \alpha^{(t)}_{K} ]$ and~$\boldsymbol{\beta}^{(t)} = [ \beta_{1}^{(t)} , \ldots , \beta_{K}^{(t)} ]$ (\cref{lin:alpha,lin:beta}, \cref{alg:RLFuzzing}),
while the parameters~$\alpha_{k}^{(0)}$ and~$\beta_{k}^{(0)}$ are initialized to~$1$ at~$t=0$ (\cref{lin:alpha-init,lin:beta-init}, \cref{alg:RLFuzzing}).
The parameter~$\boldsymbol{\alpha} - 1$ represents the number of times each~$x \in \inputcovmap$ has been covered (for all \seedinputs executed so far) by an interesting input, while~$\boldsymbol{\beta} - 1$ represents the number of times each~$x \in \inputcovmap$ was covered by input that was \emph{not} considered interesting.

Using Bayes rule, the posterior distribution~$p(\hat{\theta} | r_{k}^{(t)})$ is
\begin{equation}
\label{eqn:posterior}
\begin{aligned}
p ( \hat{\theta}^{(t)}_{k} | r_{k}^{(t)} ) &= \frac{ p(r_{k}^{(t)} | \hat{\theta}^{(t)}_{k}) p(\hat{\theta}^{(t)}_{k} ; \alpha^{(t)}_{k} , \beta^{(t)}_{k}) }{ p( r_{k}^{(t)} ) } \\
&\propto p(r_{k}^{(t)} | \hat{\theta}^{(t)}_{k}) p(\hat{\theta}^{(t)}_{k}  ; \alpha_{k}^{(t)} , \beta_{k}^{(t)} ) .
\end{aligned}    
\end{equation}

The prior distribution~$p(\hat{\theta}^{(t)}_{k}  ; \alpha_{k}^{(t)} , \beta_{k}^{(t)} )$ represents the distribution at the previous time step; i.e.,
\begin{equation}
\label{eqn:update_posterior}
p(\hat{\theta}^{(t)}_{k}  ; \alpha_{k}^{(t)} , \beta_{k}^{(t)} ) := p ( \hat{\theta}^{(t-1)}_{k} | r_{k}^{(t-1)} ) ,
\end{equation}
while the likelihood~$p(r_{k}^{(t)} | \hat{\theta}^{(t)}_{k})$ takes into account the coverage achieved at the current time step.
\Cref{eqn:update_posterior} provides a mechanism for the fuzzer to update its model to ensure it selects the ``best'' seed.
The recursive structure defined in \cref{eqn:posterior,eqn:update_posterior} means the model is dependent on previously selected inputs.
The fuzzer continuously updates the model by receiving feedback from the reward function (\cref{eqn:reward_function}) at each timestep.
This approximates the posterior distribution~$p ( \hat{\theta}^{(t)}_{k} | r_{k}^{(t)} )$, which is now the probability that a fuzzer-generated input will cover this feature in the coverage map.

\subsubsection{Improving Performance}

Directly sampling from \cref{eqn:posterior} to compute~$\boldsymbol{\theta}$ is computationally expensive.
However, using the Beta distribution (which is a \emph{conjugate prior} to the Bernoulli distribution) avoids expensive numerical computations (that are typical in Bayesian inference), leading to simpler updates after each time step.
In particular, solving \cref{eqn:posterior} with \cref{eqn:bernoulli_likelihood,eqn:beta_prior} gives the following update rule for the posterior distribution:
\begin{equation}
\label{eqn:update_posterior_distribution}
(\alpha_{k}^{(t)} , \beta_{k}^{(t)}) = (\alpha_{k}^{(t-1)} , \beta_{k}^{(t-1)} ) + (r_{k}^{(t)} , 1 - r_{k}^{(t)}) .
\end{equation}
This is the approach taken in \texttt{UpdatePosterior} (\cref{alg:RLFuzzing}):
the parameters~$\alpha^{(t)}_{k}$ and~$\beta^{(t)}_{k}$ are updated incrementally (at each time step) with each observation of success and failure, respectively.
The update rule has been illustrated in \cref{fig:toy_example_algo}, where $\alpha_{k}$ is incremented for the path of the input if the input discovered new program behavior, otherwise $\beta_{k}$ is incremented.
For example, at $t=1$, input covered lines 3 and 6 for the first time so $\alpha$ was incremented for both of these nodes, and at $t=3$, the input covered Line 6 but did not discover new code coverage, so $beta$ was incremented for Line 6.

Following these updates, \cref{eqn:posterior} is sampled by drawing from a Beta distribution with parameters~$\alpha_{k}$ and~$\beta_{k}$,
\begin{equation}
\label{eqn:sample_posterior}
\hat{\theta}^{(t)}_{k} \sim p(\theta = \hat{\theta}^{(t)}_{k} | \Tilde{r}_{k}^{(t+1)} = 1 ) = \mathrm{Beta} ( \alpha^{(t)}_{k}, \beta^{(t)}_{k} ) ,
\end{equation}
where~$\theta_{k}$---the probability of covering feature~$x \in \inputcovmap$---is dependent on the inputs seen so far.

Unfortunately, using~$\boldsymbol{\theta}$ (at \cref{lin:beta_distribution}, \cref{alg:RLFuzzing}) to select the next input will likely result in the fuzzer repeatedly selecting the same few inputs.
This is because incrementing~$\boldsymbol{\alpha}$ in \cref{eqn:update_posterior_distribution} rewards the model, ultimately skewing the probability density function (PDF) towards one and making it more likely that the fuzzer will select this input.
Similarly, incrementing~$\boldsymbol{\beta}$ penalizes the model, skewing the PDF towards zero and making it less likely that the fuzzer will select this input.
However, this has the side-effect of penalizing actions that are not selectable.
We describe this issue and how we correct for it in \cref{sec:rareness-correction}.

\subsection{Rareness Correction}
\label{sec:rareness-correction}

\cref{sec:mab-scheduler} introduced the standard MAB setting, where we assume all~$K$ actions are selectable.
However, this is not the case in practice: the fuzzer cannot select inputs corresponding to unexercised coverage map features.
Moreover, the update rule in \cref{eqn:update_posterior_distribution} penalizes rarely-covered features in~\covmap.

We use \emph{feature rareness} to penalize frequently-covered coverage map features, introducing a \emph{correction factor} to account for the update rule's penalty.
This ensures favored inputs corresponding to less-covered features in the coverage map have a greater chance of being selected, prioritizing newly-discovered and hard-to-reach code.
We apply this penalty using the chain rule on the joint probability between the reward at the next time step~$\Tilde{r}_{k}^{(t+1)}$ and the probability of the fuzzer selecting an input covering feature~$k$,~$\hat{\theta}_{k}^{(t)}$:
\begin{equation}
\label{eqn:constraint_on_rareness}
p(\Tilde{r}_{k}^{(t+1)} = 1 | \theta = \hat{\theta}^{(t)}_{k}) \propto p(\theta = \hat{\theta}^{(t)}_{k} | \Tilde{r}_{k}^{(t+1)} = 1 ) p ( \Tilde{r}_{k}^{(t+1)} = 1 ) ,
\end{equation}
where~$\Tilde{r}_{k}^{(t+1)}$ is the predicted reward at the next time step.

\cref{eqn:constraint_on_rareness} is a binary classification with dependent variables.
Here, the conditional probability with dependent variable~$\Tilde{r}_{k}^{(t+1)}=1$ is drawn from \cref{eqn:sample_posterior}, and the constraint for rareness applied by the marginal probability of~$\Tilde{r}_{k}^{(t+1)}=1$ is
\begin{equation}
\label{eqn:rareness_constraint}
\psi_{k}^{(t)} \sim p ( \Tilde{r}_{k}^{(t+1)} = 1 ) = \mathrm{Beta} \left( \alpha_{k}^{(t)} + \beta_{k}^{(t)}, \left(\alpha_{k}^{(t)} \right)^2 \right) .
\end{equation}

The conditional probability~$p(\hat{\theta}^{(t)}_{k} | \Tilde{r}_{k}^{(t+1)} = 1 )$ represents the probability action~$a_{k}$ selects an~\seedinput that will discover new behaviors.
The marginal probability~$p ( \Tilde{r}_{k}^{(t+1)} = 1 )$ applies a constraint penalizing features in the coverage map with a high hit count, thus prioritizing under-explored code.
\cref{eqn:rareness_constraint_expectation} shows this:
\begin{equation}
\label{eqn:rareness_constraint_expectation}
\phi^{(t)}_{k} = \mathbb{E} [p ( \Tilde{r}_{k}^{(t+1)} = 1 )] = \frac{ \alpha^{(t)}_{k} + \beta^{(t)}_{k} }{ \left(\alpha^{(t)}_{k} \right)^{2} + \alpha^{(t)}_{k} + \beta^{(t)}_{k} }.
\end{equation}
For~$\alpha_{k} >> \beta_{k}$, then~$\phi^{(t)}_{k} \rightarrow \frac{1}{\alpha_{k}}$; i.e., a penalty is applied if the input is selected too frequently.
Similarly, for~$\beta_{k} >> \alpha_{k},$ then~$\phi^{(t)}_{k} \rightarrow 1$; i.e., the penalty is removed if the input is infrequently chosen.
This prioritizes rare features in~\covmap .

The correction factor is self-balancing: if less-explored favored inputs fail to discover any new inputs then~$\beta_{k} \rightarrow \infty$ and~$\phi^{(t)}_{k} \rightarrow 1$.
After each time step,~$\phi_{k}$ progressively removes the penalty if the fuzzer fails to discover any new interesting behavior.
This allows~$\theta_{k}$ to dominate the seed scheduling process.
Similarly,~$\phi_{k}$ dominates if new behaviors are quickly discovered.

\subsection{Input Selection}
\label{sec:input-selection}

A fuzzer must balance exploration and exploitation when selecting the ``best'' input to fuzz.
Thompson sampling draws a value from~$K$ Beta distributions (\cref{eqn:sample_posterior}), then selects the next input to fuzz from the posterior distribution (\cref{eqn:posterior}).
Importantly, sampling a distribution generates a new distribution, enabling both exploration and exploitation.

Here, we present two approaches for selecting the next input to fuzz: input selection without and with rareness correction (\cref{sec:mab-scheduler,sec:rareness-correction}, respectively).
This enables us to evaluate rareness correction's impact on fuzzing outcomes (\cref{sec:eval-ablation}).

We express input selection \emph{without} rareness correction by sampling directly from \cref{eqn:sample_posterior}; i.e.,
\begin{equation}
\label{eqn:choose_action_wo_rareness}
a^{(t+1)} = \arg \max [ \hat{\theta}^{(t)}_{1} , \ldots , \hat{\theta}^{(t)}_{K} ] .
\end{equation}

After applying rareness correction, we express input selection as
\begin{equation}
\label{eqn:choose_action}
a^{(t+1)} = \arg \max [ \psi^{(t)}_{1} \hat{\theta}^{(t)}_{1} , \ldots , \psi^{(t)}_{K} \hat{\theta}^{(t)}_{K} ] ,
\end{equation}
where~$\psi^{(t)}_{k}$ is computed by \cref{eqn:rareness_constraint}.
The action~$a^{(t+1)}$ is then used to select the next input from the set of favored inputs (\cref{lin:select_input}, \cref{alg:RLFuzzing}).
Intuitively the parameter $a^{(t+1)}$ represents the index of an edge with the highest scheduling probability ($\theta_{k}$ or $\bar{p}$) for each time-step (e.g., for $t=4$ in \cref{fig:toy_example_algo}, $a^{(t+1)}$ will select an input that has covered Line 3 because $\bar{p}=0.31$ is the highest). 
An alternate formulation can be made by setting~$\psi^{(t)}_{k}$ to the expected value~$\phi^{(t)}_{k}$ in \cref{eqn:rareness_constraint_expectation}.

\section{Evaluation}
\label{sec:evaluation}

We evaluate \algoname over~\totalfuzztime of fuzzing, comparing it to \numcomparedfuzzers fuzzers and~\numcomparedschedulers schedulers across \numprograms programs from the Magma~\cite{Hazimeh:2020:Magma} and \fuzzbench~\cite{FuzzBench} benchmarks.
Our evaluation aims to answer the following research questions:
\begin{researchq}
\item\label{rq:bug-finding} Does \algoname improve bug discovery? (\cref{sec:eval-bug-finding})
\item\label{rq:code-coverage} Does \algoname improve code coverage? (\cref{sec:eval-code-coverage})
\item\label{rq:overheads} What are the run-time costs of \algoname? (\cref{sec:eval-overheads})
\item\label{rq:ablation} How do \algoname's design choices impact fuzzing outcomes? (\cref{sec:eval-ablation})
\end{researchq}

\subsection{Methodology}
\label{sec:methodology}

\paragraph*{Fuzzer Selection}

We evaluate three \algoname variants:
\begin{description}[style=unboxed,leftmargin=\parindent]
\item[\algoworare] No rareness correction (\cref{eqn:choose_action_wo_rareness}).

\item[\algowrare] Rareness correction via the expected value (i.e., with $\psi_{k}^{(t)} := \phi_{k}^{(t)}$ in \cref{eqn:choose_action}).

\item[\algosample] Rareness correction via sampling (\cref{eqn:choose_action}).
\end{description}
All three variants are implemented in AFL++ (v4.01a)~\cite{Fioraldi:2020:AFLPlusPlus}.
We build on AFL++---rather than AFL~\cite{Zalewski:2015:AFL}---because it incorporates state-of-the-art fuzzing improvements that \algoname can leverage.
\algowrare deterministically selects the next input by computing the model's expected value.
In contrast, \algosample probabilistically selects the next input by drawing samples from the model's distribution.

To emphasize \algoname's generality, we evaluate it against \numcomparedfuzzers fuzzers using two instrumentation schemes: LLVM compiler and QEMU binary.
While QEMU significantly reduces fuzzer iteration rates, it is important to understand fuzzer performance when source code is not available.
Importantly, this reduction is consistent across all fuzzers, so does not (dis)advantage any particular fuzzer.
We select fuzzers using LLVM to answer \cref{rq:bug-finding} and fuzzers using QEMU to answer \cref{rq:code-coverage}.
These fuzzers are:
\begin{description}[style=unboxed,leftmargin=\parindent]
\item[AFL++ (v4.01a)~\cite{Fioraldi:2020:AFLPlusPlus}]
The current state-of-the-art greybox fuzzer.
We run AFL++ with \numaflschedulers supported power schedules~\cite{Fioraldi:2020:AFLPlusPlus}:
the six \aflfast~\cite{bohme2016coverage} schedules (EXPLORE, FAST, COE, QUAD, LIN, and EXPLOIT), and AFL++'s MMOPT (increases the score for new inputs to focus on newly-discovered paths) and RARE (ignores the input's run-time and focuses on inputs covering rarely-discovered features).
We use AFL++ to answer both \cref{rq:bug-finding,rq:code-coverage}.
For \cref{rq:bug-finding} we use LLVM's link-time optimization and also fuzz with an additional ``CmpLog''-instrumented target (for logging comparison operands~\cite{Aschermann:2019:RedQueen}).

\item[\kscheduler~\cite{She:2022:KScheduler}]
Schedules inputs based on \emph{Katz centrality}~\cite{Katz:1953:Centrality} analysis of the CFG.
Katz centrality measures the ``influence'' of an input. 
This analysis helps seed scheduling by revealing the potential coverage gains from mutating a particular input.
The AFL based \kscheduler is implemented using LLVM's CFG analysis, so we use it to answer \cref{rq:bug-finding}.

\item[\tortoisefuzz~\cite{Wang:2020:TortoiseFuzz}]
Introduces three new coverage metrics for input scheduling operating on the function, loop, and basic block levels.
\tortoisefuzz uses prior information on memory operations to gain further insights in prioritizing seeds leading to memory corruption bugs.
This focus on memory corruption bugs and reliance on LLVM analyses means we use the AFL based \tortoisefuzz to answer \cref{rq:bug-finding}.

\item[\aflhier~\cite{wang2021reinforcement}]
Combines a hierarchy of coverage metrics (ranging from coarse-grained to fine-grained) and an RL-based hierarchical seed scheduler for managing clusters of inputs (preventing fine-grained coverage metrics from flooding the queue).
The AFL based \aflhier's hierarchy of coverage metrics is implemented in QEMU, and thus we use it to answer \cref{rq:code-coverage}.
\end{description}

\paragraph*{Benchmark Selection}
We evaluate these fuzzers on the Magma and \fuzzbench benchmarks.
At the time of writing, \fuzzbench's \texttt{libxml} and \texttt{libpcap} failed to download.
Thus, we omit these two targets from our evaluation.
\kscheduler also failed to construct CFGs for (and thus fuzz) \textit{php} and \textit{poppler}.

\paragraph*{Experimental Setup}
Each target is fuzzed for~\fuzztime and repeated \numtrials times to ensure statistically-sound results.
We bootstrap each target with the default seeds provided by the benchmark.
We conduct all experiments on a server with a~48-core Intel\textsuperscript{\textregistered} Xeon\textsuperscript{\textregistered} Gold 5118~\qty{2.30}{\giga\hertz} CPU,~\qty{512}{\gibi\byte} of RAM, and running Ubuntu 18.04.

\subsection{Bug Discovery (RQ~1)}
\label{sec:eval-bug-finding}

The ultimate goal of fuzzing is to find bugs.
To this end, we evaluate the LLVM-based fuzzers presented in \cref{sec:methodology} on the Magma benchmark ($\gt\totalbugfuzztime$ of fuzzing).
Magma distinguishes between bugs \emph{reached} and \emph{triggered}.
A bug is reached when ``\textit{the faulty line of code is executed}'' (i.e., control-flow constraints are met) and triggered when ``\textit{the fault condition is satisfied}'' (i.e., data-flow constraints are met).
We focus on triggering bugs (not just reaching them) and say that fuzzer/scheduler~$\mathcal{F}_1$ outperforms~$\mathcal{F}_2$ on a given bug if
\begin{inlinealph}
\item $\mathcal{F}_1$ finds the bug and~$\mathcal{F}_2$ does not, or

\item $\mathcal{F}_1$ finds the bug faster than~$\mathcal{F}_2$.
\end{inlinealph}
%


\begin{table*}
\centering
\footnotesize

\caption{Summary of Magma bug-finding results (across 10 trials).
Count = number of bugs found in a target.
Total = number of bugs found across all targets.
Best = number of times a fuzzer found the most bugs in a given target.
Unique = number of bugs found in any of the \numtrials trials.
Fastest = number of times a fuzzer found a bug first (per the restricted mean survival time~\cite{arcuri2011practical} and log-rank test~\cite{Mantel:1966:LogRankTest}).
Missed = number of times a fuzzer failed to find a bug across all \numtrials trials.
Consistency = mean number of unique bugs found per trial (i.e., total~/ unique~/~\# trials).
The best-performing fuzzer(s) for each metric  is in \myhl{hlsnsgreen}{green}.
Targets that failed to build or run with the given fuzzer are marked with \xmark.
Full results are presented in~\cref{sec:magma-survival-analysis}.
}
\label{tab:magma-bug-counts}

\begin{adjustbox}{width=\linewidth}
\begin{tabular}{l>{\itshape}l>{\ttfamily}l *{13}{r}}
\toprule
& & & \multicolumn{13}{c}{Fuzzer} \\
\cmidrule(lr){4-16}
& & & \multicolumn{8}{c}{AFL++}
& &
& \multicolumn{3}{c}{\algoname} \\
\cmidrule(lr){4-11}
\cmidrule(lr){14-16}
& \multirow{-3}{*}{\normalfont{Target}}
& \multirow{-3}{*}{\normalfont{Driver}}
& EXPLORE
& FAST
& COE
& QUAD
& LIN
& EXPLOIT
& MMOPT
& RARE
& \multirow{-2}{*}{\texttt{K-Sched}}
& \multirow{-2}{*}{Tortoise}
& \algoworare
& \algowrare
& \algosample \\
\midrule

\shaderow
\whitecell & libpng & libpng\_read\_fuzzer
& 23 & 29 & 25 & 24 & 17 & \greencell 30 & 24 & 23 & 11 & 11 & 19 & 18 & 18 \\

& libsndfile & sndfile\_fuzzer
& \greencell 70 & \greencell 70 & \greencell 70 & \greencell 70 & \greencell 70 & \greencell 70 & \greencell 70 & \greencell 70 & 28 & 20 & \greencell 70 & \greencell 70 & \greencell 70 \\

\shaderow
\whitecell & & tiff\_read\_rgba\_fuzzer
& 36 & 34 & 35 & 31 & 32 & 35 & \greencell 37 & 34 & 20 & 17 & \greencell 37 & 33 & 35 \\

\shaderow
\whitecell & \multirow{-2}{*}{libtiff} & tiffcp
& 49 & 49 & 48 & 43 & 43 & 50 & 49 & 52 & 41 & 34 & \greencell 53 & 48 & 49 \\

& & xml\_read\_memory\_fuzzer
& 31 & 34 & 34 & 30 & 34 & 30 & \greencell 40 & 31 & 10 & 10 & 30 & 34 & 35 \\

& \multirow{-2}{*}{libxml2} & xmllint
& 27 & 25 & 28 & 22 & 25 & 20 & 27 & 24 & 13 & 10 & 27 & 28 & \greencell 31 \\

\shaderow
\whitecell & lua & lua
& 10 & 10 & 10 & 10 & 6 & 9 & 9 & 10 & 10 & 9 & \greencell 13 & 11 & 11 \\

& & asn1
& 18 & \greencell 20 & 19 & 18 & 18 & \greencell 20 & \greencell 20 & 19 & 11 & 10 & \greencell 20 & \greencell 20 & \greencell 20 \\

& & client
& \greencell 10 & \greencell 10 & \greencell 10 & \greencell 10 & \greencell 10 & \greencell 10 & \greencell 10 & \greencell 10 & \greencell 10 & 3 & \greencell 10 & \greencell 10 & \greencell 10 \\

& & server
& 10 & 10 & 10 & 10 & 10 & 10 & 10 & 10 & \greencell 20 & \greencell 20 & 18 & 18 & 16 \\

& \multirow{-4}{*}{openssl} & x509
& 0 & 1 & 1 & 0 & 0 & 3 & 0 & 4 & 0 & \greencell 8 & 0 & 0 & 0 \\

\shaderow
\whitecell & php & exif
& 16 & 19 & 18 & 14 & 14 & 26 & 18 & 20 & \xmark & \greencell 30 & \greencell 30 & \greencell 30 & \greencell 30 \\

& & pdf\_fuzzer
& 32 & 27 & 27 & 25 & 25 & 31 & 30 & 30 & \xmark & 20 & 33 & 32 & \greencell 34 \\

& & pdfimages
& 35 & \greencell 42 & 40 & 29 & 25 & 32 & 38 & 32 & \xmark & 16 & 34 & 35 & 38 \\

& \multirow{-3}{*}{poppler} & pdftoppm
& 42 & 42 & \greencell 46 & 28 & 30 & 32 & \greencell 46 & 37 & \xmark & 20 & 38 & 40 & 37 \\

\shaderow
\multirow{-16}{*}{\rotatebox[origin=c]{90}{\normalfont{Count}}}
\whitecell & sqlite3 & sqlite3\_fuzz
& 45 & \greencell 50 & 38 & 29 & 49 & 33 & 42 & 39 & 2 & 0 & 36 & 41 & 43 \\
\midrule

\shaderow
\multicolumn{3}{l}{Total}
                           & 454 & 472 & 459 & 393 & 408 & 441 & 471 & 445
                           & 176
                           & 238
                           & 468
                           & 468
                           & \greencell 477 \\
\multicolumn{3}{l}{Best}
                           & 2 & 5 & 3 & 2 & 2 & 4 & 6 & 2
                           & 2
                           & 3
                           & \greencell 7
                           & 4
                           & 6 \\
\shaderow
\multicolumn{3}{l}{Unique}
                           & 63 & 63 & \greencell 65 & 54 & 61 & 59 & 63 & 60
                           & 24
                           & 29
                           & 62
                           & 63
                           & \greencell 65 \\
\multicolumn{3}{l}{Fastest}
                           & \greencell 11 & 10 & 4 & 6 & 8 & 7 & 9 & 4
                           & 1
                           & 3
                           & 7
                           & 5
                           & \greencell 11 \\
\shaderow
\multicolumn{3}{l}{Missed}
                           & 15 & 15 & \greencell 13 & 24 & 17 & 19 & 15 & 18
                           & 54
                           & 49
                           & 16
                           & 15
                           & \greencell 13 \\
\midrule
\multicolumn{3}{l}{Consistency}
                           & 0.72 & 0.75 & 0.71 & 0.73 & 0.67 & 0.75 & 0.75 & 0.74
                           & 0.73
                           & \greencell 0.82
                           & 0.75
                           & 0.73
                           & 0.73 \\
\bottomrule
\end{tabular}
\end{adjustbox}
\end{table*}

\Cref{tab:magma-bug-counts} shows that \algosample was the best-performing scheduler across four out of five (\qty{80}{\percent}) of our bug-finding metrics (``total'', ``unique'', ``fastest'', and ``missed''), while \algoworare scored the highest on the remaining metric (``best'').
%
Of the \numaflschedulers AFL++ schedulers, FAST found the most bugs (``total''~=~472) and was the second-best performer across the ``best'', ``unique'', ``fastest'', and ``missed'' metrics (after MMOPT, COE, EXPLORE, and COE, respectively).
These results reinforce the AFL++ developers' decision to make FAST the default scheduler.
Notably, the EXPLOIT scheduler (the original AFL's scheduler) was one of the worst performers (e.g., it was the third worst performer in missed bugs).

COE found more unique bugs (65) than \algoworare (62) and \algowrare (63).
However, the higher ``total'' results across the \algoname variants suggest that \algoname produces more-consistent bug-finding results.
We reinforce this result with a ``consistency'' metric, defined as the total number of bugs divided by the number of unique bugs averaged across all \numtrials trials.
Based on this metric, \algoworare outperformed or performed as well as all AFL++ schedulers.
That is, the number of bugs discovered by \algoname remains consistent as the number of trials decreases.

All three \algoname variants outperformed \kscheduler and \tortoisefuzz---two state-of-the-art schedulers---across four of the five (\qty{80}{\percent}) bug-finding metrics (``total'', ``best'', ``unique'', and ``missed'').
\algosample (the best-performing \algoname variant) outperformed or performed as well as \kscheduler on~11 out of~12 drivers (we exclude \textit{php} and \textit{poppler} because they failed to build).
Similarly, \algosample outperformed or performed as well as \tortoisefuzz on~14 out of~16 drivers.
We examine these results in the following sections.

\subsubsection{\kscheduler Comparison}
\label{sec:kscheduler-bug-compare}

Of the~35 bugs found by \algosample or \kscheduler, the former:
\begin{inlineroman}
\item outperformed \kscheduler on~27 bugs;

\item performed as well as \kscheduler on six bugs; and

\item was outperformed by \kscheduler on two bugs.
\end{inlineroman}
Notably, \kscheduler failed to discover~16 of the~27 (\qty{59}{\percent}) bugs where \algoname outperformed \kscheduler.
Moreover, \kscheduler was, on average,~\appx$50\times$ slower at finding the remaining ten bugs.
We attribute these results to \kscheduler prioritizing ``exploration'' over ``exploitation''.
Prior work on directed greybox fuzzing~\cite{Bohme:2017:AFLGo,Wang:2020:DGFSoK,Zheng:2022:FishFuzz} has shown that fuzzers must both ``explore'' interesting code and ``exploit'' specific data-flow conditions to trigger bugs.\footnote{This is an unfortunate overloading of terms, and should not be confused with a MAB's exploration and exploitation phases.}
Concentrating on CFG expansion means that \kscheduler does not focus on this exploitation phase, potentially harming bug discovery.



\paragraph*{Finding: TIF009}

One exception to these results is TIF009, which \kscheduler found after~\qty{3.29}{\hour} (mean time over \numtrials trials).
In comparison, \algoworare, \algowrare, and \algosample found it after \qtylist{14.31;33.37;33.77}{\hour}, respectively (4--10$\times$ slower).
TIF009 (CVE-2019-7663~\cite{CVE-2019-7663}) is a \texttt{NULL} pointer dereference located in \texttt{TIFFWriteDirectoryTagTransferfunction}.
This vulnerable function is reachable via \texttt{TIFFWriteDirectory} when the \texttt{TransferFunction} field is set in a TIFF directory entry~\cite{TIFFSpec}, and the bug is triggered when the transfer function pointers are \texttt{NULL}.

\kscheduler and \algoname trigger the bug within~\qtyrange{3}{120}{\second} of reaching it, suggesting the bug does not require satisfying complex data-flow constraints (which pure mutational fuzzers---e.g., AFL++---may have difficulty satisfying).
Moreover, multiple inputs from the initial seed set reach \texttt{TIFFWriteDirectory} without \emph{any} mutation.
Ultimately, \kscheduler's centrality score led it to the vulnerable function faster because it prioritized inputs that explored \texttt{TIFFWriteDirectory}, which had a relatively high centrality score (0.288).
In comparison, neighboring CFG nodes had significantly lower centrality scores (0.047; while the median centrality score was~0.22).
The relatively-simple data-flow constraints (a struct field set to \texttt{NULL}) meant that \kscheduler's prioritization of exploration over exploitation was an advantage.

\subsubsection{\tortoisefuzz Comparison}
\label{sec:tortoise-bug-compare}

37 of the~55 (\qty{67}{\percent}) bugs discovered across all~13 fuzzers/schedulers were memory safety bugs (e.g., stack/heap buffer overflow/over-read, \texttt{NULL} pointer dereference).
Thus, we expected \tortoisefuzz to demonstrate superior results in discovering these bugs (due to its design targeting memory safety bugs).
However, \tortoisefuzz was outperformed by the other fuzzers/schedulers (in particular, \algoname): it failed to trigger~20 of the~37 (\qty{54}{\percent}) memory safety bugs in \emph{any} trial, and was slower at triggering another eight.
\tortoisefuzz was outperformed by \algoworare, \algowrare, and \algosample on~51,~53, and~58 bugs, respectively.

Notably \tortoisefuzz and \kscheduler use a different set of heuristics based on the assumption that inputs reaching specific target sites are more likely to induce new program behaviors.
\tortoisefuzz prioritizes memory-sensitive bugs allowing it to detect several memory-sensitive bugs earlier such as \texttt{TIF009}, \texttt{SSL009}, and \texttt{PDF010}.
In contrast, \algoname heuristics are derived from the assumption that
\begin{inlinealph}
\item information derived from previous test cases can be used to improve input scheduling in the future, and
\item it is beneficial to spend resources on paths less explored to gain more information about the program to improve input scheduling.
\end{inlinealph}
A potential improvement is to extend the model to be a directed greybox fuzzing where they have a different set of assumptions where the scheduler should prioritize specific sites such as memory-sensitive functions which are more likely to induce new program behaviors.

\paragraph*{Finding: \textit{sqlite3}}

\tortoisefuzz failed to find any bugs in \textit{sqlite3}.
In contrast, \algoworare found seven bugs, while \algowrare and \algosample found eight.
This was due to the limited coverage expanded by \tortoisefuzz: it achieved only~\appx\qty{7}{\percent} line coverage, while \algosample achieved~$\gt\qty{50}{\percent}$ line coverage.
We attribute this low coverage to \tortoisefuzz's iteration rate: only~\qty{58}{\input\per\second}.
In contrast, \algoname achieved an iteration rate~$\gt\qty{450}{\input\per\second}$.
This reinforces the importance of reducing run-time overheads to maximize throughput.

\begin{hlbox}{Result 1}
\algosample was the (equal) best performer across the most (four out of five) bug-finding metrics: it found the most unique bugs, was the fastest, and missed the fewest bugs.
\algoworare outperformed in the ``best'' metric.
\end{hlbox}

\subsection{Code Coverage (RQ~2)}
\label{sec:eval-code-coverage}

Bugs are sparse, making it challenging to evaluate fuzzers fairly using bug-centric metrics.
Fuzzer evaluations commonly use code coverage as a proxy for evaluating fuzzer performance (a bug cannot be found in code never executed).
We repeat this practice here.

We compare coverage using QEMU binary instrumentation on~\numfuzzbenchtargets \fuzzbench targets ($\gt\totalcovfuzztime$ of fuzzing).
Comparisons are made across two measures: final coverage and coverage area under curve (AUC).
Coverage is measured using Clang's source-based coverage instrumentation~\cite{clang:2022:SrcCov}.
We use AUC because we found several targets had maximized coverage before the end of a~\qty{72}{\hour} trial. 
A higher AUC indicates that the fuzzer uncovers behaviors at a faster rate, which is important if the length of a fuzzing campaign is constrained.
We use the Mann-Whitney~$U$ test to determine statistical significance~\cite{Klees:2018:EvaluatingFuzzing}.

\begin{table*}
\centering
\footnotesize
\caption{\fuzzbench coverage, presented as mean coverage with~\qty{95}{\percent} bootstrap~CI.
The best-performing fuzzer(s) for each target (per the Mann-Whitney~$U$ test) is in \myhl{hlsnsgreen}{green} (larger is better).
``Best'' is the number of targets a fuzzer achieved the highest coverage.
}
\label{tab:fuzzbench-cov}


\begin{adjustbox}{width=\linewidth}
\begin{tabular}{>{\ttfamily}l *{12}{r}}
\toprule
& \multicolumn{12}{c}{Fuzzer} \\
\cmidrule(lr){2-13}
& \multicolumn{8}{c}{AFL++}
&
& \multicolumn{3}{c}{\algoname} \\
\cmidrule(lr){2-9}
\cmidrule(lr){11-13}
\multirow{-3}{*}{\normalfont{Target}}
& EXPLORE
& FAST
& COE
& QUAD
& LIN
& EXPLOIT
& MMOPT
& RARE
& \multirow{-2}{*}{\aflhier}
& \algoworare
& \algowrare
& \algosample \\
\midrule

\shaderow
& \num{5884.70} & \num{5767.60} & \greencell \num{5998.50} & \num{5514.40} & \num{5794.30} & \greencell \num{6069.50} & \greencell \num{5944.10} & \num{5935.50} & \greencell \num{6113.20} & \num{5918.20} & \num{6001.70} & \greencell \num{6069.20} \\
\shaderow
\multirow{-2}{*}{bloaty}
& \ci{109.18} & \ci{97.83} & \greencell \ci{74.35} & \ci{131.59} & \ci{108.91} & \greencell \ci{61.48} & \greencell \ci{141.67} & \ci{88.13} & \greencell \ci{126.46} & \ci{41.84} & \ci{47.54} & \greencell \ci{47.18} \\

& \num{18968.10} & \num{16778.90} & \greencell \num{19375.40} & \num{18738.90} & \num{18660.50} & \num{18204.40} & \num{19164.40} & \num{18712.20} & \num{13370.10} & \num{16795.20} & \num{16776.30} & \num{16786.00} \\
\multirow{-2}{*}{curl}                 
& \ci{46.26} & \ci{35.11} & \greencell \ci{45.64} & \ci{149.82} & \ci{64.69} & \ci{148.61} & \ci{72.55} & \ci{88.77} & \ci{9.31} & \ci{124.08} & \ci{53.54} & \ci{79.30} \\

\shaderow
& \num{14396.00} & \num{14585.80} & \greencell \num{16419.70} & \num{13836.00} & \num{14246.40} & \num{15609.50} & \num{15053.80} & \num{15364.90} & \greencell \num{14598.80} & \greencell \num{15860.40} & \greencell \num{16018.10} & \num{15734.60} \\
\shaderow
\multirow{-2}{*}{freetype2}              
& \ci{197.46} & \ci{356.16} & \greencell \ci{181.27} & \ci{272.73} & \ci{204.66} & \ci{275.23} & \ci{399.48} & \ci{201.14} & \greencell \ci{924.03} & \greencell \ci{171.62} & \greencell \ci{219.93} & \ci{176.92} \\

& \greencell \num{7642.40} & \num{7550.50} & \greencell \num{7777.10} & \num{7333.60} & \num{7222.30} & \num{7470.00} & \greencell \num{7640.00} & \num{7560.30} & \num{6616.10} & \greencell \num{7727.60} & \num{7577.60} & \greencell \num{7663.00} \\
\multirow{-2}{*}{harfbuzz}
& \greencell \ci{118.41} & \ci{104.96} & \greencell \ci{58.23} & \ci{144.77} & \ci{115.52} & \ci{71.30} & \greencell \ci{101.47} & \ci{70.59} & \ci{317.23} & \greencell \ci{31.45} & \ci{35.36} & \greencell \ci{24.29} \\

\shaderow                                     
& \greencell \num{639.00} & \greencell \num{639.00} & \greencell \num{639.00} & \greencell \num{639.00} & \greencell \num{639.00} & \num{636.00} & \greencell \num{639.00} & \greencell \num{639.00} & \num{545.10} & \greencell \num{639.00} & \greencell \num{639.00} & \greencell \num{639.00} \\
\shaderow
\multirow{-2}{*}{jsoncpp}                    
& \greencell \ci{0.00} & \greencell \ci{0.00} & \greencell \ci{0.00} & \greencell \ci{0.00} & \greencell \ci{0.00} & \ci{0.00} & \greencell \ci{0.00} & \greencell \ci{0.00} & \ci{56.91} & \greencell \ci{0.00} & \greencell \ci{0.00} & \greencell \ci{0.00} \\

& \num{1227.20} & \num{1223.80} & \num{1317.10} & \num{1235.80} & \num{1370.10} & \num{1262.00} & \num{1224.70} & \num{1235.30} & \greencell \num{2293.22} & \num{1254.70} & \num{1233.40} & \num{1251.40} \\
\multirow{-2}{*}{lcms}
& \ci{0.58} & \ci{1.18} & \ci{17.45} & \ci{8.95} & \ci{136.16} & \ci{16.67} & \ci{1.83} & \ci{6.69} & \greencell \ci{135.79} & \ci{11.71} & \ci{2.09} & \ci{13.38} \\

\shaderow
& \num{3561.10} & \num{3656.60} & \greencell \num{3745.90} & \num{3451.40} & \num{3442.40} & \num{3140.70} & \num{3749.80} & \num{3057.90} & \num{2909.60} & \greencell \num{3764.40} & \num{3497.00} & \num{3517.60} \\
\shaderow
\multirow{-2}{*}{libjpeg-turbo}
& \ci{55.74} & \ci{49.42} & \greencell \ci{41.05} & \ci{44.76} & \ci{21.84} & \ci{36.43} & \ci{31.54} & \ci{11.70} & \ci{75.03} & \greencell \ci{29.95} & \ci{47.71} & \ci{54.57} \\

& \greencell \num{2089.40} & \greencell \num{2084.10} & \greencell \num{2084.50} & \num{2076.10} & \num{2082.40} & \greencell \num{2091.40} & \greencell \num{2097.40} & \num{2056.20} & \num{1619.00} & \num{2079.20} & \num{2074.30} & \num{2064.10} \\
\multirow{-2}{*}{libpng}
& \greencell \ci{4.80} & \greencell \ci{5.07} & \greencell \ci{7.94} & \ci{5.29} & \ci{6.32} & \greencell \ci{4.53} & \greencell \ci{5.72} & \ci{7.82} & \ci{32.11} & \ci{6.11} & \ci{6.13} & \ci{7.27} \\

\shaderow
& \num{7698.00} & \num{7642.30} & \greencell \num{8147.80} & \num{7659.90} & \num{7701.80} & \num{7749.20} & \num{7652.20} & \num{7693.10} & \num{7467.10} & \num{8025.70} & \num{7652.70} & \num{7631.20} \\
\shaderow
\multirow{-2}{*}{mbedtls}
& \ci{10.83} & \ci{19.18} & \greencell \ci{17.13} & \ci{13.89} & \ci{13.86} & \ci{13.00} & \ci{5.68} & \ci{16.09} & \ci{100.39} & \ci{26.42} & \ci{21.38} & \ci{24.86} \\

& \num{13731.20} & \greencell \num{13745.80} & \greencell \num{13750.80} & \num{13720.10} & \num{13730.30} & \greencell \num{13754.30} & \num{13738.80} & \num{13728.60} & \num{13695.30} & \num{13738.40} & \num{13731.80} & \num{13735.80} \\
\multirow{-2}{*}{openssl}                                         
& \ci{3.34} & \greencell \ci{3.13} & \greencell \ci{3.72} & \ci{2.15} & \ci{4.85} & \greencell \ci{2.10} & \ci{4.11} & \ci{4.80} & \ci{4.64} & \ci{4.33} & \ci{3.71} & \ci{4.42} \\

\shaderow                 
& \num{5643.20} & \greencell \num{5764.80} & \num{5703.10} & \num{5364.10} & \num{5390.20} & \greencell \num{5706.60} & \greencell \num{5740.70} & \num{5750.50} & \num{5424.22} & \greencell \num{5802.70} & \num{5594.80} & \num{5617.40} \\
\shaderow
\multirow{-2}{*}{openthread}
& \ci{38.42} & \greencell \ci{20.37} & \ci{31.87} & \ci{42.34} & \ci{59.08} & \greencell \ci{15.50} & \greencell \ci{29.01} & \ci{32.94} & \ci{157.82} & \greencell \ci{18.32} & \ci{79.15} & \ci{55.67} \\

& \num{42342.10} & \num{42459.60} & \num{42961.10} & \num{41627.00} & \num{41505.80} & \num{41907.40} & \num{42364.50} & \num{42344.90} & \num{41117.30} & \greencell \num{44736.60} & \num{42190.00} & \num{44198.10} \\
\multirow{-2}{*}{php}                             
& \ci{106.95} & \ci{100.18} & \ci{61.12} & \ci{61.70} & \ci{38.58} & \ci{109.88} & \ci{136.80} & \ci{99.31} & \ci{534.67} & \greencell \ci{70.01} & \ci{143.45} & \ci{73.78} \\

\shaderow                 
& \num{5610.70} & \num{5699.10} & \greencell \num{6579.70} & \num{5132.90} & \num{5112.60} & \num{5310.30} & \num{5807.10} & \num{5129.00} & \num{2320.80} & \greencell \num{6466.50} & \greencell \num{6092.50} & \num{6132.00} \\
\shaderow
\multirow{-2}{*}{proj4}
& \ci{141.28} & \ci{76.92} & \greencell \ci{91.73} & \ci{80.43} & \ci{54.38} & \ci{132.76} & \ci{121.24} & \ci{86.67} & \ci{369.80} & \greencell \ci{106.58} & \greencell \ci{205.82} & \ci{67.97} \\

& \num{3506.10} & \num{3498.50} & \greencell \num{3518.00} & \num{3486.00} & \num{3504.00} & \num{3475.70} & \num{3502.60} & \num{3502.70} & \num{3055.30} & \greencell \num{3530.10} & \num{3486.70} & \num{3484.50} \\
\multirow{-2}{*}{re2}
& \ci{3.59} & \ci{4.15} & \greencell \ci{4.18} & \ci{7.05} & \ci{1.69} & \ci{4.83} & \ci{5.71} & \ci{6.87} & \ci{22.87} & \greencell \ci{5.40} & \ci{7.58} & \ci{4.90} \\

\shaderow
& \num{21191.10} & \num{21651.90} & \greencell \num{23210.00} & \num{21281.40} & \num{21558.00} & \greencell \num{23197.10} & \num{22097.90} & \greencell \num{22479.50} & \num{19041.60} & \greencell \num{23301.90} & \num{22028.80} & \num{22021.20} \\
\shaderow
\multirow{-2}{*}{sqlite3}
& \ci{151.57} & \ci{306.67} & \greencell \ci{366.46} & \ci{113.28} & \ci{159.46} & \greencell \ci{386.04} & \ci{194.40} & \greencell \ci{127.28} & \ci{177.51} & \greencell \ci{577.27} & \ci{78.95} & \ci{143.47} \\

& \num{637.60} & \greencell \num{639.40} & \greencell \num{640.00} & \greencell \num{640.00} & \greencell \num{638.20} & \num{633.40} & \greencell \num{638.80} & \greencell \num{639.40} & \num{613.56} & \greencell \num{640.00} & \greencell \num{640.00} & \greencell \num{640.00} \\
\multirow{-2}{*}{systemd}
& \ci{0.92} & \greencell \ci{0.56} & \greencell \ci{0.00} & \greencell \ci{0.00} & \greencell \ci{1.21} & \ci{1.57} & \greencell \ci{1.14} & \greencell \ci{0.57} & \ci{13.21} & \greencell \ci{0.00} & \greencell \ci{0.00} & \greencell \ci{0.00} \\

\shaderow
& \num{2098.40} & \num{1991.70} & \greencell \num{2140.40} & \num{2030.90} & \num{2061.70} & \num{2043.70} & \num{2087.10} & \num{1986.90} & \num{1549.00} & \greencell \num{2156.30} & \num{2073.20} & \num{2062.70} \\
\shaderow
\multirow{-2}{*}{vorbis}
& \ci{16.27} & \ci{39.81} & \greencell \ci{8.86} & \ci{26.69} & \ci{32.15} & \ci{17.80} & \ci{18.38} & \ci{23.53} & \ci{0.00} & \greencell \ci{4.42} & \ci{8.68} & \ci{22.25} \\

& \num{1769.80} & \num{1694.40} & \greencell \num{1841.30} & \num{1773.10} & \num{1744.60} & \num{1731.40} & \num{1775.10} & \num{1678.00} & \num{1498.56} & \greencell \num{1860.90} & \num{1717.60} & \num{1699.10} \\
\multirow{-2}{*}{woff2}
& \ci{14.61} & \ci{13.88} & \greencell \ci{9.47} & \ci{7.82} & \ci{14.93} & \ci{5.95} & \ci{6.69} & \ci{8.61} & \ci{49.29} & \greencell \ci{9.41} & \ci{5.69} & \ci{7.14} \\

\shaderow
& \num{945.40} & \num{942.20} & \greencell \num{953.80} & \num{932.80} & \num{932.20} & \num{940.20} & \num{947.20} & \greencell \num{948.50} & \num{920.30} & \greencell \num{960.10} & \greencell \num{954.90} & \greencell \num{959.40} \\
\shaderow
\multirow{-2}{*}{zlib}
& \ci{4.12} & \ci{5.24} & \greencell \ci{3.91} & \ci{3.71} & \ci{3.42} & \ci{4.81} & \ci{5.06} & \greencell \ci{5.02} & \ci{3.69} & \greencell \ci{1.65} & \greencell \ci{4.67} & \greencell \ci{3.49} \\

\midrule
\normalfont{Best} & 3 & 5 & \greencell 16 & 2 & 2 & 5 & 6 & 4 & 3 & 13 & 5 & 5 \\

\bottomrule
\end{tabular}
\end{adjustbox}
\end{table*}

\begin{table*}
\centering
\footnotesize
\caption{\fuzzbench AUC, presented as mean AUC with~\qty{95}{\percent} bootstrap~CI.
The best-performing fuzzer(s) for each target (per the Mann-Whitney~$U$ test) is in \myhl{hlsnsgreen}{green} (larger is better).
``Best'' is the number of times a fuzzer achieved the highest AUC for the evaluated targets.
}
\label{tab:fuzzbench-auc}


\begin{adjustbox}{width=\linewidth}
\begin{tabular}{>{\ttfamily}l *{12}{r}}
\toprule
& \multicolumn{12}{c}{Fuzzer} \\
\cmidrule(lr){2-13}
& \multicolumn{8}{c}{AFL++}
&
& \multicolumn{3}{c}{\algoname} \\
\cmidrule(lr){2-9}
\cmidrule(lr){11-13}
\multirow{-3}{*}{\normalfont{Target}}
& EXPLORE
& FAST
& COE
& QUAD
& LIN
& EXPLOIT
& MMOPT
& RARE
& \multirow{-2}{*}{\aflhier}
& \algoworare
& \algowrare
& \algosample \\
\midrule

\shaderow
& \greencell \num{1855.49} & \num{1798.74} & \greencell \num{1881.75} & \num{1739.14} & \num{1832.82} & \greencell \num{1901.46} & \greencell \num{1860.15} & \greencell \num{1882.21} & \num{720.66} & \greencell \num{1876.27} & \greencell \num{1903.64} & \greencell \num{1917.34} \\
\shaderow
\multirow{-2}{*}{bloaty}
& \greencell \ci{36.81} & \ci{29.02} & \greencell \ci{21.23} & \ci{37.05} & \ci{32.55} & \greencell \ci{18.05} & \greencell \ci{39.57} & \greencell \ci{26.01} & \ci{88.80} & \greencell \ci{12.40} & \greencell \ci{12.70} & \greencell \ci{12.21} \\

& \num{5935.13} & \num{5297.44} & \greencell \num{6128.42} & \num{5876.09} & \num{5845.46} & \num{5685.59} & \num{5981.73} & \num{5899.21} & \num{22.31} & \num{5270.48} & \num{5284.60} & \num{5278.60} \\
\multirow{-2}{*}{curl}                 
& \ci{32.79} & \ci{8.34} & \greencell \ci{21.96} & \ci{49.73} & \ci{32.48} & \ci{42.23} & \ci{11.66} & \ci{16.57} & \ci{4.41} & \ci{31.77} & \ci{15.11} & \ci{23.71} \\

\shaderow
& \num{4485.76} & \num{4524.82} & \greencell \num{5026.24} & \num{4316.93} & \num{4433.65} & \num{4794.04} & \num{4617.88} & \num{4691.47} & \num{1596.44} & \greencell \num{4892.82} & \greencell \num{4911.98} & \num{4821.05} \\
\shaderow
\multirow{-2}{*}{freetype2}              
& \ci{63.39} & \ci{102.24} & \greencell \ci{53.00} & \ci{83.98} & \ci{52.09} & \ci{71.91} & \ci{108.91} & \ci{37.17} & \ci{349.75} & \greencell \ci{58.15} & \greencell \ci{46.05} & \ci{42.39} \\

& \greencell \num{2367.14} & \num{2326.66} & \greencell \num{2407.32} & \num{2270.86} & \num{2237.21} & \num{2292.35} & \greencell \num{2358.83} & \num{2340.57} & \num{1317.19} & \greencell \num{2425.86} & \greencell \num{2372.57} & \num{2396.13} \\
\multirow{-2}{*}{harfbuzz}
& \greencell \ci{34.39} & \ci{31.08} & \greencell \ci{16.32} & \ci{41.14} & \ci{30.87} & \ci{21.88} & \greencell \ci{30.49} & \ci{21.70} & \ci{211.41} & \greencell \ci{11.06} & \greencell \ci{13.43} & \ci{7.54} \\

\shaderow                   
& \num{203.63} & \num{203.23} & \greencell \num{203.71} & \num{203.54} & \num{203.69} & \num{202.66} & \num{203.46} & \num{203.49} & \num{80.35} & \num{202.70} & \num{203.39} & \num{203.27} \\                  
\shaderow
\multirow{-2}{*}{jsoncpp}                    
& \ci{0.07} & \ci{0.17} & \greencell \ci{0.05} & \ci{0.02} & \ci{0.01} & \ci{0.02} & \ci{0.11} & \ci{0.01} & \ci{15.07} & \ci{0.28} & \ci{0.17} & \ci{0.15} \\

& \num{389.70} & \num{387.03} & \num{393.74} & \num{390.46} &  \num{406.59} & \num{392.60} & \num{388.93} & \num{390.42} & \greencell \num{572.34} & \num{394.88} & \num{389.15} & \num{390.45} \\
\multirow{-2}{*}{lcms}
& \ci{0.33} & \ci{0.57} & \ci{0.81} & \ci{0.96} & \ci{15.81} & \ci{1.47} & \ci{0.49} & \ci{0.46} & \greencell \ci{48.40} & \ci{1.99} & \ci{0.57} & \ci{1.55} \\

\shaderow
& \num{1085.49} & \num{1096.15} & \greencell \num{1160.83} & \num{1071.70} & \num{1072.38} & \num{973.90} & \num{1121.60} & \num{954.81} & \num{750.88} & \greencell \num{1173.16} & \num{1064.67} & \num{1067.75} \\
\shaderow
\multirow{-2}{*}{libjpeg-turbo}
& \ci{9.90} & \ci{11.33} & \greencell \ci{11.89} & \ci{8.03} & \ci{3.32} & \ci{4.95} & \ci{12.41} & \ci{6.06} & \ci{38.03} & \greencell \ci{11.36} & \ci{15.36} & \ci{16.61} \\

& \greencell \num{663.27} & \greencell \num{660.20} & \greencell \num{663.96} & \num{657.04} & \num{660.73} & \num{661.04} & \greencell \num{665.76} & \num{650.26} & \num{442.56} & \num{658.86} & \num{657.18} & \num{652.64} \\
\multirow{-2}{*}{libpng}
& \greencell \ci{1.60} & \greencell \ci{1.99} & \greencell \ci{2.50} & \ci{2.20} & \ci{2.07} & \ci{1.43} & \greencell \ci{1.78} & \ci{2.72} & \ci{28.84} & \ci{2.12} & \ci{1.97} & \ci{2.12} \\

\shaderow
& \num{2436.35} & \num{2419.82} & \greencell \num{2541.22} & \num{2436.52} & \num{2446.75} & \num{2455.33} & \num{2420.35} & \num{2436.75} & \num{1941.63} & \num{2471.21} & \num{2428.81} & \num{2418.06} \\
\shaderow
\multirow{-2}{*}{mbedtls}
& \ci{2.24} & \ci{5.30} & \greencell \ci{5.61} & \ci{3.85} & \ci{3.71} & \ci{3.24} & \ci{3.02} & \ci{3.06} & \ci{231.06} & \ci{9.09} & \ci{6.12} & \ci{7.74} \\

& \num{4369.59} & \num{4370.03} & \greencell \num{4382.95} & \num{4372.61} & \num{4376.14} & \num{4378.27} & \num{4373.62} & \num{4374.11} & \num{2414.84} & \num{4355.91} & \num{4377.09} & \num{4368.88} \\
\multirow{-2}{*}{openssl}                                         
& \ci{3.07} & \ci{4.42} & \greencell \ci{1.09} & \ci{0.94} & \ci{1.51} & \ci{0.70} & \ci{1.99} & \ci{2.01} & \ci{513.73} & \ci{7.69} & \ci{1.14} & \ci{4.03} \\

\shaderow         
& \num{1748.75} & \num{1762.49} & \num{1791.13} & \num{1668.19} & \num{1660.39} & \num{1746.66} & \num{1790.89} & \num{1784.31} & \num{1254.70} & \greencell \num{1825.09} & \num{1729.62} & \num{1713.15} \\        
\shaderow
\multirow{-2}{*}{openthread}
& \ci{17.22} & \ci{12.70} & \ci{11.93} & \ci{13.06} & \ci{19.64} & \ci{8.34} & \ci{10.25} & \ci{6.85} & \ci{180.57} & \greencell \ci{6.25} & \ci{21.97} & \ci{16.16} \\

& \num{13390.60} & \num{13388.88} & \num{13584.21} & \num{13194.36} & \num{13177.75} & \num{13183.46} & \num{13389.12} & \num{13337.44} & \num{1151.67} & \greencell \num{14016.82} & \num{13414.13} & \greencell \num{14016.68} \\
\multirow{-2}{*}{php}                             
& \ci{31.39} & \ci{32.36} & \ci{23.81} & \ci{9.33} & \ci{9.90} & \ci{25.63} & \ci{36.87} & \ci{23.16} & \ci{280.18} & \greencell \ci{19.42} & \ci{38.98} & \greencell \ci{26.92} \\

\shaderow          
& \num{1647.87} & \num{1670.72} & \greencell \num{2020.40} & \num{1491.32} & \num{1467.89} & \num{1426.01} & \num{1671.35} & \num{1327.46} & \num{436.16} & \greencell \num{1977.17} & \num{1770.04} & \num{1785.75} \\       
\shaderow
\multirow{-2}{*}{proj4}
& \ci{43.23} & \ci{28.16} & \greencell \ci{28.88} & \ci{30.43} & \ci{23.12} & \ci{41.85} & \ci{46.43} & \ci{30.11} & \ci{76.83} & \greencell \ci{39.66} & \ci{66.98} & \ci{25.78} \\

& \num{1107.15} & \num{1102.90} & \num{1116.69} & \num{1097.35} & \num{1099.13} & \num{1082.91} & \num{1103.85} & \num{1097.64} & \num{719.14} & \greencell \num{1121.36} & \num{1100.35} & \num{1095.72} \\
\multirow{-2}{*}{re2}
& \ci{1.24} & \ci{1.06} & \ci{0.94} & \ci{1.67} & \ci{1.17} & \ci{2.76} & \ci{1.47} & \ci{1.79} & \ci{32.05} & \greencell \ci{1.38} & \ci{2.12} & \ci{2.18} \\

\shaderow
& \num{6546.16} & \num{6549.26} & \greencell \num{6966.93} & \num{6659.31} & \num{6684.83} & \greencell \num{6776.79} & \num{6770.59} & \greencell \num{6917.85} & \num{3163.01} & \greencell \num{7032.32} & \num{6862.27} & \num{6815.92} \\
\shaderow
\multirow{-2}{*}{sqlite3}
& \ci{37.64} & \ci{84.99} & \greencell \ci{70.91} & \ci{23.54} & \ci{42.71} & \greencell \ci{62.16} & \ci{31.57} & \greencell \ci{23.47} & \ci{228.67} & \greencell \ci{78.15} & \ci{20.75} & \ci{31.30} \\

& \num{202.70} & \num{202.93} & \greencell \num{204.00} & \num{202.95} & \num{202.97} & \num{200.68} & \num{202.86} & \num{202.53} & \num{148.28} & \greencell \num{203.38} & \num{203.29} & \num{203.16} \\
\multirow{-2}{*}{systemd}
& \ci{0.37} & \ci{0.20} & \greencell \ci{0.02} & \ci{0.25} & \ci{0.41} & \ci{0.33} & \ci{0.45} & \ci{0.28} & \ci{26.43} & \greencell \ci{0.28} & \ci{0.16} & \ci{0.26} \\

\shaderow
& \num{621.37} & \num{597.84} & \num{655.33} & \num{613.50} & \num{618.62} & \num{627.95} & \num{624.02} & \num{616.59} & \num{2.07} & \greencell \num{669.33} & \num{625.59} & \num{625.32} \\
\shaderow
\multirow{-2}{*}{vorbis}
& \ci{4.06} & \ci{3.77} & \ci{2.96} & \ci{4.78} & \ci{4.97} & \ci{4.83} & \ci{5.24} & \ci{5.06} & \ci{0.22} & \greencell \ci{1.99} & \ci{4.40} & \ci{4.44} \\

& \num{540.76} & \num{528.67} & \num{564.95} & \num{549.08} & \num{543.36} & \num{540.87} & \num{541.73} & \num{526.76} & \num{197.68} & \greencell \num{580.63} & \num{538.27} & \num{533.78} \\
\multirow{-2}{*}{woff2}
& \ci{1.79} & \ci{2.04} & \ci{2.12} & \ci{1.01} & \ci{3.23} & \ci{1.03} & \ci{1.71} & \ci{1.75} & \ci{73.95} & \greencell \ci{2.47} & \ci{1.04} & \ci{1.61} \\

\shaderow
& \greencell \num{299.20} & \greencell \num{298.56} & \greencell \num{301.50} & \num{296.07} & \num{295.10} & \num{291.20} & \greencell \num{300.56} & \greencell \num{299.10} & \num{273.63} & \greencell \num{302.06} & \greencell \num{301.07} & \greencell \num{302.36} \\
\shaderow
\multirow{-2}{*}{zlib}
& \greencell \ci{1.05} & \greencell \ci{1.48} & \greencell \ci{1.22} & \ci{0.77} & \ci{0.50} & \ci{0.75} & \greencell \ci{1.34} & \greencell \ci{1.35} & \ci{10.12} & \greencell \ci{1.34} & \greencell \ci{1.26} & \greencell \ci{1.04} \\

\midrule
\normalfont{Best} & 4 & 2 & \greencell 13 & 0 & 0 & 2 & 4 & 3 & 1 & \greencell 13 & 4 & 3 \\

\bottomrule
\end{tabular}
\end{adjustbox}
\end{table*}


Per \cref{tab:fuzzbench-cov}, AFL++'s COE achieved the (equal) highest coverage on~16 of the \numfuzzbenchtargets targets (\qty{84}{\percent}).
Of the \algoname variants, \algoworare achieved the highest coverage on~13 targets (\qty{68}{\percent}) and was the next best performer after COE.
\algowrare and \algosample performed similarly (equal best on five targets).
\algowrare's smaller~\qty{95}{\percent} bootstrap~CI indicates less variance (across trials); we attribute this to \algowrare's deterministic approach for rareness correction.
\aflhier was the third worst performing fuzzers (only beating QUAD and LIN, and tying with EXPLORE).
Curiously, it was the best performer on \texttt{lcms}, achieving twice as much coverage as the next best fuzzer.

COE was again the (equal) best performer for AUC (\cref{tab:fuzzbench-auc}).
However, this time it also tied with \algoworare.
Following \citet{Bohme:2022:FuzzerReliability}, we use Cohen's kappa~\cite{Cohen:1969:Kappa} to measure the agreement between total coverage and AUC of the best-performing fuzzer(s).
We found the results in \cref{tab:fuzzbench-cov,tab:fuzzbench-auc} are in weak agreement ($\kappa = 0.56$).
Outliers including \texttt{jsoncpp}, \texttt{openssl}, and \texttt{systemd} contributed to this weak agreement.
For example, \algoworare outperformed \algowrare and \algosample on \texttt{openssl} in \cref{tab:fuzzbench-cov}, while the opposite is true in \cref{tab:fuzzbench-auc}.
These results were due to inaccurate AUC measurements: coverage measurements occur at~\qty{15}{\minute} intervals, and all fuzzers had saturated within the first~\qty{15}{\minute}.

\begin{hlbox}{Result 2}
COE achieved the (equal) highest coverage on~16 of the~19 targets (\qty{84}{\percent}).
\algoworare tied with COE when using AUC.
\end{hlbox}

\subsection{Scheduler Overheads (RQ~3)}
\label{sec:eval-overheads}

Scheduler overhead impacts a fuzzer's iteration rate.
Prior work~\cite{Xu:2017:OSPrimitives,Herrera:2021:SeedSelection} has shown that increased iteration rates lead to improved fuzzing outcomes.
Here we investigate the scheduler's impact on iteration rates and fuzzing outcomes. 
%
To measure this impact, we:
\begin{inlineroman}
\item instrument the scheduler to compute run-time overhead, recording the time the fuzzer spends updating the queue and selecting an input to fuzz;

\item count the number of times the queue is updated (i.e., the number of times the fuzzer performs ``filtering and favoring''); and

\item examine the iteration rate reported by the fuzzer.
\end{inlineroman}
We adopt the process used by \citet{She:2022:KScheduler} and run our instrumentation across~\numfuzzbenchtargets \fuzzbench targets for~\qty{24}{\hour}, repeating this experiment \numtrials times to minimize variance.

\begin{table*}
\centering
\footnotesize

\caption{Scheduler overheads and iteration rates with~\qty{95}{\percent} bootstrap~CI.
Update count = number of times the queue is updated in a single trial (geometric mean).
Update time = time spent (ms) on each queue update (arithmetic mean).
Update variance = how much the queue update time varies (ms\textsuperscript{2}) in a single trial (arith.\ mean).
Overhead = total time (s) the fuzzer spends selecting an input to fuzz in a trial (arith.\ mean).
Iteration rate = number of inputs executed per second (arith.\ mean).
}
\label{tab:overhead-summary}

\begin{adjustbox}{width=\linewidth}
\begin{tabular}{l *{12}{r}}
\toprule
& \multicolumn{12}{c}{Fuzzer} \\
\cmidrule(lr){2-13}
& \multicolumn{8}{c}{AFL++}
&
& \multicolumn{3}{c}{\algoname} \\
\cmidrule(lr){2-9}
\cmidrule(lr){11-13}
& EXPLORE
& FAST
& COE
& QUAD
& LIN
& EXPLOIT
& MMOPT
& RARE
& \multirow{-2}{*}{\aflhier}
& \algoworare
& \algowrare
& \algosample \\
\midrule

\shaderow
& \num{269.97} & \num{873.86} & \num{283.12} & \num{328.12} & \num{590.47} & \num{80.03} & \num{281.25} & \num{124.42} & \num{88.84} & \num{672.38} & \num{649.44} & \num{635.84} \\
\shaderow
\multirow{-2}{*}{Update count (\#)}
& \ci{44.62} & \ci{156.92} & \ci{46.29} & \ci{57.84} & \ci{135.07} & \ci{7.91} & \ci{46.43} & \ci{17.73} & \ci{17.17} & \ci{87.68} & \ci{86.14} & \ci{81.63} \\

& \num{14.96} & \num{9.29} & \num{9.52} & \num{15.17} & \num{16.77} & \num{29.88} & \num{15.62} & \num{39.77} & \num{106.55} & \num{41.76} & \num{42.05} & \num{79.83} \\
\multirow{-2}{*}{Update time (ms)}
& \ci{3.47} & \ci{1.32} & \ci{1.30} & \ci{3.26} & \ci{7.29} & \ci{7.25} & \ci{2.83} & \ci{7.99} & \ci{23.59} & \ci{0.09} & \ci{0.11} & \ci{0.23} \\

\shaderow
& \num{0.40} & \num{0.01} & \num{0.00} & \num{0.09} & \num{0.12} & \num{0.11} & \num{0.04} & \num{0.52} & \num{0.75} & \num{0.00} & \num{0.00} & \num{0.00} \\
\shaderow
\multirow{-2}{*}{Update variance (ms\textsuperscript{2})}
& \ci{0.37} & \ci{0.00} & \ci{0.00} & \ci{0.04} & \ci{0.10} & \ci{0.06} & \ci{0.02} & \ci{0.32} & \ci{0.51} & \ci{0.00} & \ci{0.00} & \ci{0.00} \\

& \num{57.84} & \num{85.35} & \num{49.34} & \num{58.18} & \num{62.00} & \num{49.18} & \num{56.38} & \num{64.13} & \num{207.68} & \num{2488.36} & \num{2419.26} & \num{4443.65} \\
\multirow{-2}{*}{Overhead (s)}
& \ci{9.32} & \ci{11.90} & \ci{7.66} & \ci{9.37} & \ci{9.23} & \ci{7.02} & \ci{8.71} & \ci{10.14} & \ci{47.91} & \ci{319.63} & \ci{317.35} & \ci{559.37} \\

\midrule
\shaderow
& \num{83.52} & \num{210.19} & \num{85.06} & \num{89.26} & \num{87.61} & \num{99.86} & \num{89.42} & \num{58.52} & \num{84.84} & \num{96.79} & \num{94.46} & \num{93.60} \\
\shaderow
\multirow{-2}{*}{Iteration rate (inputs/s)}
& \ci{13.33} &\ci{39.23} & \ci{12.20} & \ci{16.73} & \ci{15.89} & \ci{16.94} & \ci{13.30} & \ci{9.79} & \ci{17.28} & \ci{16.75} & \ci{16.36} & \ci{16.67} \\

\bottomrule
\end{tabular}
\end{adjustbox}
\end{table*}


\Cref{tab:overhead-summary} shows that, despite an orders-of-magnitude increase in overhead, \algoname achieves iteration rates comparable to (and, in most cases, higher than) the \numaflschedulers AFL++ schedulers.
%
Of the \numaflschedulers AFL++ schedulers, FAST had the highest overhead (\qty{85}{\second}), which we attribute to the high number of queue updates (873 over a single~\qty{24}{\hour} trial).
However, this again did not impact iteration rates; FAST's iteration rate of~\qty{210}{\input\per\second} was the highest.
The \texttt{libpng} and \texttt{zlib} targets dominated this result, achieving iteration rates over~\qtylist{500;700}{\input\per\second}, respectively.

Unlike \algoname, \aflhier's relatively high update time (\qty{106}{\milli\second}) and overhead (\qty{3}{\minute}) appeared to impact its iteration rate, which was the third lowest.
We attribute these results to the hierarchical tree structure \aflhier uses to abstract the queue.

\subsubsection{Scalability}
\label{sec:scheduler-scalability}

Queue update times should remain constant (per trial) as the fuzzer expands coverage and the queue grows.
This is particularly important for large fuzzing campaigns (e.g., Google's OSS-Fuzz~\cite{Serebryany:2017:OSSFuzz}), where the queue can grow to tens of thousands of inputs.
To this end, we investigate a scheduler's scalability by examining how much queue update times vary (across a single~\qty{24}{\hour} trial).
\Cref{tab:overhead-summary} shows these results under ``update variance''.

Like queue update time, queue update time variance is negligible.
\algoname has effectively no variance, making it ideal for long-running fuzzing campaigns.
\aflhier and AFL++'s RARE have the highest variance.
This is unsurprising; RARE focuses on queue entries that hit rare coverage map elements, requiring additional computation to find those rare elements, while operations on \aflhier's hierarchical tree structure have~$\mathcal{O}(n)$ complexity.

\begin{hlbox}{Result 3}
\algoname maintains high iteration rates, despite increased scheduling overheads.
\algoname is also scalable: queue update times remain constant over fuzzing campaigns.
\end{hlbox}

\subsection{Ablation Study (RQ~4)}
\label{sec:eval-ablation}

We undertake an ablation study to understand better the impacts the individual \algoname components have on our results.
We study:
\begin{inlineroman}
\item AFL++ FAST;

\item \algoworare, replacing the FAST scheduler with a MAB; and

\item \algosample, adding a correction for rareness.
\end{inlineroman}
We select FAST because it is AFL++'s default scheduler, and \algosample because it generally outperforms \algowrare (\cref{rq:bug-finding,rq:code-coverage}).

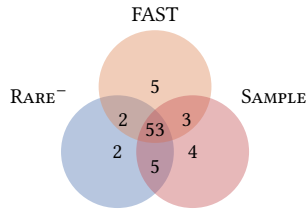
\begin{figure}
\centering
\small

\begin{tikzpicture}[
  thick,
  set/.style = {circle, minimum size=1.5cm, opacity=0.4, text opacity=1}
]

\node (algoworare) [set, fill=snsblue, label={135:\algoworare}] {2};
\node (fast) at (60:1cm) [set, fill=snsorange, label={FAST}] {5};
\node (algosample) at (0:1cm) [set, fill=snsred, label={45:\algosample}] {4};
 
\node at (barycentric cs:algoworare=1,fast=1) [left] {2};
\node at (barycentric cs:algoworare=1,algosample=1) [below] {5};
\node at (barycentric cs:fast=1,algosample=1) [right] {3};
\node at (barycentric cs:algoworare=1,fast=1,algosample=1) [] {53};

\end{tikzpicture}

\caption{Venn diagram of unique bug count between \aflfast, \algoworare, and \algosample (74 unique bugs found across all drivers).}
\label{fig:bug-venn}

\end{figure}

\Cref{fig:bug-venn} visualizes the unique bug counts from \cref{tab:magma-bug-counts}.
FAST found~63 unique bugs, of which five were only found by FAST.
Replacing FAST with \algoworare decreased the number of unique bugs to~62.
However, \algoworare found two bugs missed by the other two schedulers.
When adding rareness correction (via \algosample) to \algoworare, the number of bugs increased to~65.
Moreover, four of these~65 bugs were only found by \algosample, and were found
\begin{inlinealph}
\item late into the fuzzing process (after~\appx\qty{60}{\hour}), and

\item by only a small number of fuzzers/schedulers,
\end{inlinealph}
suggesting they are difficult to trigger.

\algosample's bug-finding performance improved towards the end of the fuzzing campaign (here, after~\qty{45}{\hour} of fuzzing), triggering nine bugs that were missed by FAST.
Three of these bugs (\texttt{XML001}, \texttt{XML002}, and \texttt{PDF011}) are buffer overflow bugs in functions that are frequently covered by FAST.
FAST assigns higher priority to an input depending on the number of times it covers an edge, while penalizing inputs that cover paths with high frequency~\cite{bohme2016coverage}.
This is a limitation of FAST, as a high-frequency path does not necessarily mean a lower chance of discovering new program behavior.
\algoname does not apply a penalty to high-frequency paths if previously scheduled inputs have discovered new program behaviors.

Code coverage increased when replacing the FAST scheduler with \algoworare's MAB-based scheduler.
Notably, \algoworare achieved the (equal) highest coverage on most \fuzzbench targets (\cref{tab:fuzzbench-cov,tab:fuzzbench-auc}).
In particular, \texttt{harfbuzz} was the only target where FAST achieved the most coverage and did not tie with \algoworare (\cref{tab:fuzzbench-cov}).
This is in contrast to bug finding, where \algosample outperformed \algoworare (except for the ``best'' metric in \cref{tab:magma-bug-counts}).
We attribute this to \algoworare's single objective (in the RL algorithm) to maximize code coverage.
In contrast, \algosample uses a multi-objective optimization that balances exploring rare paths and maximizing code coverage.

Iteration rates decreased by~\appx\qty{50}{\percent} when replacing FAST with \algoworare (\cref{tab:overhead-summary}).
However, this decrease was also true of the other seven AFL++ schedulers (and was sometimes even more pronounced; up to~\qty{72}{\percent}).
\algosample increased the scheduler overhead from~\qty{40}{\minute} to~\qty{74}{\minute} (an~\qty{84}{\percent} increase).
We attribute this increase to costs associated with sampling from the Beta distribution twice (\cref{eqn:choose_action}).
In contrast, \algoworare only samples from it once.
However, this has negligible impact on iteration rates: a reduction of~\appx\qty{1}{\input\per\second}.
Similarly, the variance across queue update times remains zero when replacing \algoworare with \algosample, demonstrating its scalability.

\paragraph*{To \algosample or not to \algosample?}

Our ablation study focused on \algosample, which probabilistically samples for rareness correction.
However, it is worth revisiting \algowrare---which deterministically computes an expected value for rareness correction---to understand what impact the probabilistic approach has on fuzzing outcomes.
\algosample outperformed \algowrare across all bug-finding metrics (\cref{tab:magma-bug-counts}).
\algosample also slightly outperformed \algowrare (by a single target) on total coverage (\cref{tab:fuzzbench-cov}) and performed the same on coverage~AUC (\cref{tab:fuzzbench-auc}).
While probabilistically sampling resulted in higher scheduler overhead, this had a negligible impact on the fuzzer's iteration rate (\cref{tab:overhead-summary}).
In most applications, the probabilistic approach should outperform a deterministic approach.
Thus, we recommend \algosample for general use based on our results.

\begin{hlbox}{Result 4}
Accounting for rare coverage elements (with \algosample) improves bug-finding performance (over \algoworare).
However, \algoworare's single objective of maximizing coverage leads to higher coverage.
Both improve upon FAST.
\end{hlbox}

\section{Related Work}
\label{sec:related-work}

\Citet{woo2013scheduling} were one of the first to formulate fuzzing as a MAB.
Their work focused on \emph{blackbox} fuzzing, where there is no coverage feedback to guide the fuzzer.
\Citet{woo2013scheduling} proposed using the exp3 algorithm~\cite{seldin2013evaluation}, with the fuzzer rewarded based on the number of unique bugs found.
However, bugs are sparse, so assigning a reward based on their discovery is impractical.
In contrast, \aflfast~\cite{bohme2016coverage} focused on \emph{greybox} fuzzing, introducing the power schedule for seed scheduling (see \cref{sec:seed-scheduling-background}).
\entropic~\cite{bohme2020boosting} expanded this with an entropy-based power schedule for prioritizing \emph{information gain}.

Prior work has applied MAB to greybox fuzzing.
\citet{Siddharth:2018:ThompsonFuzzing,Koike:2022:SLOPT} use a MAB over the fuzzer's mutation operators, while EcoFuzz~\cite{yue2020ecofuzz}, \aflhier~\cite{wang2021reinforcement}, and MobFuzz~\cite{Zhang:2022:MobFuzz} use a MAB for seed scheduling.

EcoFuzz~\cite{yue2020ecofuzz} eschews a traditional MAB for an \emph{adversarial bandit}.
While an adversarial bandit removes assumptions about the bandits' probability distributions, this generality requires more hyperparameters to tune.
In particular, EcoFuzz has two hyperparameters---exploration and decay---and uses entropy to compute the probability of an input's (potential) information gain.

\aflhier~\cite{wang2021reinforcement} combines a multiplayer MAB with a hierarchical tree structure to balance exploration/exploitation across coverage metrics.
Like EcoFuzz, \aflhier has two hyperparameters (exploration and decay).
However, unlike EcoFuzz, \aflhier rewards rare code paths, encouraging the scheduler to select inputs that cover less-explored paths.

Finally, MobFuzz~\cite{Zhang:2022:MobFuzz} uses a multi-objective optimization formulated as a multiplayer MAB, maximizing three objectives: execution speed, memory consumption, and the length of memory comparisons.
MobFuzz uses a weighted average of these three objectives to compute the reward for each input.

\section{Conclusions}
\label{sec:conclusions}

We present \algoname, an RL-based fuzzer seed scheduler.
Unlike prior RL-based schedulers, \algoname does not require hyperparameter tuning.
We empirically demonstrate \algoname's effectiveness over~\totalfuzztime of fuzzing.
On Magma, \algoname found the most bugs and missed the fewest.
On \fuzzbench, it was the fastest at expanding coverage on the most targets.
\algoname also maintains consistently-high iteration rates (even as the queue grows).
Given our results, we recommend the adoption of our \algosample variant.
Interestingly, our results also show that AFL++'s default FAST scheduler was generally outperformed by the MMOPT scheduler (across both bug-finding and coverage-expansion metrics).
We encourage others to adopt and build upon \algoname, available at \artifacturl.

\begin{acks}
This work was supported by the Defence Science and Technology Group Next Generation Technologies Fund (Cyber) program via the \emph{KRONOS} Data61 Collaborative Research Project.
\end{acks}

\bibliographystyle{ACM-Reference-Format}
\bibliography{references}

\appendix

\section{Magma Survival Analysis}
\label{sec:magma-survival-analysis}

Following prior work~\cite{arcuri2011practical,Wagner:2017:ElasticProgramTransform,Hazimeh:2020:Magma,Herrera:2021:SeedSelection,Herrera:2022:datAFLow}, we model bug finding using survival analysis.
This allows us to reason about censored data; i.e., the case where a fuzzer does not find a bug.
\Cref{tab:magma-survival} presents the restricted mean survival time (RMST) of a given bug; i.e., the mean time the bug ``survives'' being discovered by a fuzzer across \numtrials repeated~\fuzztime campaigns.
Lower RMSTs imply a fuzzer finds a bug ``faster'', while a smaller confidence interval (CI) means the bug is found more consistently.
Applying the log-rank test~\cite{Mantel:1966:LogRankTest} under the null hypothesis that two fuzzers share the same survival function allows us to statistically compare survival times.
Thus, two fuzzers have statistically equivalent bug survival times if the log-rank test's~$\pvalue > 0.05$.
The survival analysis results in \cref{tab:magma-survival} augment those presented in the main paper.

\begin{table*}
\centering
\footnotesize

\caption{Magma bugs triggered, presented as the restricted mean survival time (RMST; in hours) with~\qty{95}{\percent} bootstrap~CI.
Bugs never found by a particular fuzzer have an RMST of~$\top$ (to distinguish bugs with a~\fuzztime RMST).
Targets that fail to build with a given fuzzer are marked with~\xmark.
The best-performing fuzzer (fuzzers if the bug survival times are statistically equivalent per the log-rank test) for each bug is highlighted in \myhl{hlsnsgreen}{green} (smaller is better).}
\label{tab:magma-survival}

\begin{adjustbox}{width=\linewidth}
\begin{tabular}{>{\itshape}l>{\ttfamily}ll *{13}{r}}
\toprule
& & & \multicolumn{13}{c}{Fuzzer} \\
\cmidrule(lr){4-16}
& & & \multicolumn{8}{c}{AFL++} & & & \multicolumn{3}{c}{\algoname} \\
\cmidrule(lr){4-11}
\cmidrule(lr){14-16}
\multirow{-3}{*}{\normalfont{Target}}
& \multirow{-3}{*}{\normalfont{Driver}}
& \multirow{-3}{*}{Bug}
& EXPLORE
& FAST
& COE
& QUAD
& LIN
& EXPLOIT
& MMOPT
& RARE
& \multirow{-2}{*}{\texttt{K-Sched}}
& \multirow{-2}{*}{Tortoise}
& \algoworare
& \algowrare
& \algosample \\
\midrule
&
&
& \greencell \num{71.51}
& $\top$
& $\top$
& $\top$
& $\top$
& \greencell \num{70.53}
& $\top$
& $\top$
& $\top$
& $\top$
& $\top$
& $\top$
& $\top$ \\
&
& \multirow{-2}{*}{PNG001}
& \greencell \ci{1.67}
& 
& 
& 
& 
& \greencell \ci{5.00}
& 
& 
& 
& 
& 
& 
&  \\
\cmidrule(lr){3-3}
\cmidrule(lr){4-16}

&
&
& \greencell \num{14.40}
& \greencell \num{0.01}
& \greencell \num{0.01}
& \greencell \num{7.21}
& \num{28.80}
& \greencell \num{0.01}
& \greencell \num{7.21}
& \greencell \num{7.21}
& \greencell \num{0.01}
& \num{0.01}
& \greencell \num{0.01}
& \greencell \num{0.01}
& \greencell \num{0.01} \\
&
& \multirow{-2}{*}{PNG003}
& \greencell \ci{25.24}
& \greencell \ci{0.00}
& \greencell \ci{0.01}
& \greencell \ci{14.11}
& \ci{30.92}
& \greencell \ci{0.01}
& \greencell \ci{18.93}
& \greencell \ci{17.28}
& \greencell \ci{0.01}
& \ci{0.01}
& \greencell \ci{0.01}
& \greencell \ci{0.01}
& \greencell \ci{0.01} \\
\cmidrule(lr){3-3}
\cmidrule(lr){4-16}

&
&
& \num{14.45}
& \num{0.08}
& \greencell \num{0.04}
& \greencell \num{7.24}
& \num{28.83}
& \greencell \num{0.05}
& \greencell \num{7.25}
& \num{7.26}
& $\top$
& $\top$
& $\top$
& $\top$
& $\top$ \\
&
& \multirow{-2}{*}{PNG006}
& \ci{17.84}
& \ci{0.05}
& \greencell \ci{0.02}
& \greencell \ci{12.76}
& \ci{24.43}
& \greencell \ci{0.03}
& \greencell \ci{12.75}
& \ci{12.75}
& 
& 
& 
& 
&  \\
\cmidrule(lr){3-3}
\cmidrule(lr){4-16}

&
&
& \num{39.28}
& \greencell \num{35.25}
& \num{51.15}
& \greencell \num{38.13}
& \num{47.51}
& \greencell \num{26.85}
& \num{42.04}
& \num{52.84}
& \num{68.36}
& \num{70.22}
& \greencell \num{28.31}
& \greencell \num{30.63}
& \greencell \num{28.02} \\
\multirow{-9}{*}{libpng} & \multirow{-9}{*}{libpng\_read\_fuzzer} &  \multirow{-2}{*}{PNG007}
& \ci{19.37}
& \greencell \ci{13.41}
& \ci{19.63}
& \greencell \ci{21.00}
& \ci{20.87}
& \greencell \ci{14.18}
& \ci{19.27}
& \ci{18.18}
& \ci{12.35}
& \ci{6.03}
& \greencell \ci{15.21}
& \greencell \ci{14.04}
& \greencell \ci{16.30} \\
\cmidrule(lr){1-1}
\cmidrule(lr){2-2}
\cmidrule(lr){3-3}
\cmidrule(lr){4-16}
&
&
& \num{0.64}
& \num{0.41}
& \num{0.46}
& \num{1.29}
& \num{1.43}
& \num{2.46}
& \num{0.56}
& \num{0.45}
& \num{34.02}
& $\top$
& \greencell \num{0.24}
& \greencell \num{0.21}
& \num{0.32} \\
&
& \multirow{-2}{*}{SND001}
& \ci{0.24}
& \ci{0.11}
& \ci{0.21}
& \ci{0.52}
& \ci{0.31}
& \ci{1.64}
& \ci{0.36}
& \ci{0.17}
& \ci{0.53}
& 
& \greencell \ci{0.08}
& \greencell \ci{0.11}
& \ci{0.08} \\
\cmidrule(lr){3-3}
\cmidrule(lr){4-16}

&
&
& \num{0.97}
& \num{0.78}
& \num{1.09}
& \num{3.92}
& \num{2.88}
& \num{6.57}
& \num{1.51}
& \num{1.02}
& $\top$
& \num{2.82}
& \greencell \num{0.41}
& \greencell \num{0.55}
& \greencell \num{0.48} \\
&
& \multirow{-2}{*}{SND005}
& \ci{0.27}
& \ci{0.32}
& \ci{0.42}
& \ci{1.48}
& \ci{1.07}
& \ci{3.59}
& \ci{0.68}
& \ci{0.43}
& 
& \ci{1.20}
& \greencell \ci{0.10}
& \greencell \ci{0.24}
& \greencell \ci{0.13} \\
\cmidrule(lr){3-3}
\cmidrule(lr){4-16}

&
&
& \num{1.11}
& \greencell \num{1.10}
& \greencell \num{0.85}
& \num{0.98}
& \num{5.69}
& \num{6.36}
& \num{1.00}
& \greencell \num{0.34}
& \num{68.24}
& $\top$
& \greencell \num{0.40}
& \greencell \num{0.45}
& \greencell \num{0.36} \\
&
& \multirow{-2}{*}{SND006}
& \ci{0.86}
& \greencell \ci{1.23}
& \greencell \ci{0.51}
& \ci{0.46}
& \ci{7.29}
& \ci{2.68}
& \ci{0.44}
& \greencell \ci{0.14}
& \ci{12.76}
& 
& \greencell \ci{0.14}
& \greencell \ci{0.19}
& \greencell \ci{0.08} \\
\cmidrule(lr){3-3}
\cmidrule(lr){4-16}

&
&
& \greencell \num{0.70}
& \greencell \num{0.85}
& \greencell \num{0.46}
& \num{1.27}
& \num{1.57}
& \num{2.86}
& \num{1.27}
& \greencell \num{0.66}
& \num{56.23}
& $\top$
& \greencell \num{0.60}
& \greencell \num{0.80}
& \greencell \num{0.79} \\
&
& \multirow{-2}{*}{SND007}
& \greencell \ci{0.32}
& \greencell \ci{0.30}
& \greencell \ci{0.27}
& \ci{0.53}
& \ci{0.61}
& \ci{1.42}
& \ci{0.56}
& \greencell \ci{0.27}
& \ci{15.46}
& 
& \greencell \ci{0.26}
& \greencell \ci{0.18}
& \greencell \ci{0.31} \\
\cmidrule(lr){3-3}
\cmidrule(lr){4-16}

&
&
& \greencell \num{0.34}
& \greencell \num{0.47}
& \greencell \num{0.57}
& \greencell \num{0.89}
& \num{1.67}
& \greencell \num{0.59}
& \greencell \num{0.57}
& \greencell \num{0.74}
& \num{1.94}
& \greencell \num{0.67}
& \greencell \num{1.35}
& \greencell \num{0.36}
& \greencell \num{0.34} \\
&
& \multirow{-2}{*}{SND017}
& \greencell \ci{0.19}
& \greencell \ci{0.31}
& \greencell \ci{0.23}
& \greencell \ci{0.69}
& \ci{1.19}
& \greencell \ci{0.15}
& \greencell \ci{0.20}
& \greencell \ci{0.41}
& \ci{0.12}
& \greencell \ci{0.13}
& \greencell \ci{0.90}
& \greencell \ci{0.31}
& \greencell \ci{0.22} \\
\cmidrule(lr){3-3}
\cmidrule(lr){4-16}

&
&
& \greencell \num{0.75}
& \greencell \num{0.80}
& \greencell \num{1.06}
& \num{1.40}
& \num{2.18}
& \num{2.03}
& \greencell \num{1.12}
& \greencell \num{1.14}
& $\top$
& $\top$
& \num{2.96}
& \num{3.36}
& \num{2.63} \\
&
& \multirow{-2}{*}{SND020}
& \greencell \ci{0.30}
& \greencell \ci{0.29}
& \greencell \ci{0.21}
& \ci{0.49}
& \ci{0.83}
& \ci{0.74}
& \greencell \ci{0.25}
& \greencell \ci{0.27}
& 
& 
& \ci{0.93}
& \ci{1.52}
& \ci{0.96} \\
\cmidrule(lr){3-3}
\cmidrule(lr){4-16}

&
&
& \greencell \num{0.59}
& \greencell \num{0.38}
& \greencell \num{0.30}
& \num{0.98}
& \num{0.93}
& \num{2.62}
& \num{0.97}
& \greencell \num{0.34}
& \num{60.41}
& $\top$
& \greencell \num{0.38}
& \greencell \num{0.45}
& \greencell \num{0.35} \\
\multirow{-17}{*}{libsndfile} & \multirow{-17}{*}{sndfile\_fuzzer} &  \multirow{-2}{*}{SND024}
& \greencell \ci{0.27}
& \greencell \ci{0.27}
& \greencell \ci{0.14}
& \ci{0.46}
& \ci{0.37}
& \ci{1.27}
& \ci{0.43}
& \greencell \ci{0.14}
& \ci{15.52}
& 
& \greencell \ci{0.15}
& \greencell \ci{0.19}
& \greencell \ci{0.08} \\
\cmidrule(lr){1-1}
\cmidrule(lr){2-2}
\cmidrule(lr){3-3}
\cmidrule(lr){4-16}
&
&
& \greencell \num{60.02}
& \num{60.46}
& \greencell \num{60.19}
& \num{65.84}
& \num{66.72}
& \num{62.47}
& \greencell \num{56.93}
& \greencell \num{58.95}
& $\top$
& $\top$
& \greencell \num{58.99}
& \num{66.96}
& \num{64.17} \\
&
& \multirow{-2}{*}{TIF002}
& \greencell \ci{15.10}
& \ci{18.66}
& \greencell \ci{10.33}
& \ci{20.91}
& \ci{11.50}
& \ci{14.48}
& \greencell \ci{15.73}
& \greencell \ci{17.79}
& 
& 
& \greencell \ci{13.38}
& \ci{8.80}
& \ci{12.92} \\
\cmidrule(lr){3-3}
\cmidrule(lr){4-16}

&
&
& \num{0.07}
& \num{0.08}
& \num{0.04}
& \num{0.12}
& \num{0.06}
& \num{0.05}
& \greencell \num{0.03}
& \num{0.04}
& \num{1.66}
& \num{4.45}
& \greencell \num{0.03}
& \num{0.04}
& \greencell \num{0.02} \\
&
& \multirow{-2}{*}{TIF007}
& \ci{0.04}
& \ci{0.03}
& \ci{0.02}
& \ci{0.14}
& \ci{0.03}
& \ci{0.02}
& \greencell \ci{0.02}
& \ci{0.03}
& \ci{0.40}
& \ci{1.58}
& \greencell \ci{0.02}
& \ci{0.03}
& \greencell \ci{0.01} \\
\cmidrule(lr){3-3}
\cmidrule(lr){4-16}

&
&
& \greencell \num{67.16}
& \greencell \num{64.98}
& $\top$
& $\top$
& $\top$
& \greencell \num{66.81}
& \greencell \num{63.17}
& \greencell \num{67.89}
& $\top$
& $\top$
& \greencell \num{66.63}
& $\top$
& \greencell \num{64.90} \\
&
& \multirow{-2}{*}{TIF008}
& \greencell \ci{9.80}
& \greencell \ci{23.84}
& 
& 
& 
& \greencell \ci{11.22}
& \greencell \ci{17.41}
& \greencell \ci{13.95}
& 
& 
& \greencell \ci{14.58}
& 
& \greencell \ci{14.50} \\
\cmidrule(lr){3-3}
\cmidrule(lr){4-16}

&
&
& \greencell \num{1.52}
& \greencell \num{1.92}
& \greencell \num{1.25}
& \num{3.05}
& \num{1.75}
& \greencell \num{1.44}
& \greencell \num{1.35}
& \num{1.84}
& \num{2.42}
& \num{51.10}
& \greencell \num{1.37}
& \greencell \num{0.97}
& \greencell \num{0.90} \\
&
& \multirow{-2}{*}{TIF012}
& \greencell \ci{0.56}
& \greencell \ci{1.01}
& \greencell \ci{0.34}
& \ci{1.04}
& \ci{0.35}
& \greencell \ci{0.72}
& \greencell \ci{0.36}
& \ci{0.49}
& \ci{0.54}
& \ci{18.80}
& \greencell \ci{0.66}
& \greencell \ci{0.34}
& \greencell \ci{0.39} \\
\cmidrule(lr){3-3}
\cmidrule(lr){4-16}

&
&
& \num{5.63}
& \greencell \num{2.72}
& \num{4.17}
& \greencell \num{4.12}
& \greencell \num{3.11}
& \greencell \num{2.49}
& \greencell \num{3.68}
& \greencell \num{1.59}
& $\top$
& \num{64.30}
& \greencell \num{2.15}
& \num{3.85}
& \greencell \num{2.04} \\
& \multirow{-12}{*}{tiff\_read\_rgba\_fuzzer} &  \multirow{-2}{*}{TIF014}
& \ci{2.44}
& \greencell \ci{1.17}
& \ci{1.69}
& \greencell \ci{2.89}
& \greencell \ci{2.39}
& \greencell \ci{1.27}
& \greencell \ci{2.52}
& \greencell \ci{0.65}
& 
& \ci{19.23}
& \greencell \ci{1.41}
& \ci{2.27}
& \greencell \ci{0.98} \\
\cmidrule(lr){2-2}
\cmidrule(lr){3-3}
\cmidrule(lr){4-16}

&
&
& $\top$
& \greencell \num{68.29}
& $\top$
& $\top$
& $\top$
& \greencell \num{69.71}
& \greencell \num{70.72}
& \greencell \num{66.34}
& $\top$
& $\top$
& \greencell \num{65.47}
& $\top$
& \greencell \num{66.71} \\
&
& \multirow{-2}{*}{TIF002}
& 
& \greencell \ci{12.58}
& 
& 
& 
& \greencell \ci{7.78}
& \greencell \ci{4.35}
& \greencell \ci{10.97}
& 
& 
& \greencell \ci{15.84}
& 
& \greencell \ci{10.45} \\
\cmidrule(lr){3-3}
\cmidrule(lr){4-16}

&
&
& \greencell \num{69.44}
& \greencell \num{65.94}
& \greencell \num{65.84}
& $\top$
& \greencell \num{61.04}
& \greencell \num{66.74}
& $\top$
& $\top$
& $\top$
& $\top$
& \greencell \num{68.74}
& $\top$
& $\top$ \\
&
& \multirow{-2}{*}{TIF005}
& \greencell \ci{8.68}
& \greencell \ci{20.57}
& \greencell \ci{20.90}
& 
& \greencell \ci{22.01}
& \greencell \ci{10.31}
& 
& 
& 
& 
& \greencell \ci{11.05}
& 
&  \\
\cmidrule(lr){3-3}
\cmidrule(lr){4-16}

&
&
& \greencell \num{22.19}
& \greencell \num{22.62}
& \greencell \num{13.46}
& \num{51.00}
& \num{46.21}
& \num{31.89}
& \greencell \num{16.42}
& \greencell \num{12.05}
& \num{64.89}
& \num{41.90}
& \greencell \num{14.92}
& \greencell \num{20.82}
& \greencell \num{20.53} \\
&
& \multirow{-2}{*}{TIF006}
& \greencell \ci{8.76}
& \greencell \ci{13.97}
& \greencell \ci{5.32}
& \ci{17.15}
& \ci{22.09}
& \ci{14.87}
& \greencell \ci{13.61}
& \greencell \ci{4.82}
& \ci{24.15}
& \ci{17.22}
& \greencell \ci{7.82}
& \greencell \ci{12.40}
& \greencell \ci{9.97} \\
\cmidrule(lr){3-3}
\cmidrule(lr){4-16}

&
&
& \greencell \num{0.05}
& \greencell \num{0.06}
& \num{0.17}
& \num{0.14}
& \greencell \num{0.05}
& \num{0.07}
& \num{0.05}
& \greencell \num{0.05}
& \num{0.23}
& \num{9.52}
& \greencell \num{0.04}
& \greencell \num{0.04}
& \greencell \num{0.03} \\
&
& \multirow{-2}{*}{TIF007}
& \greencell \ci{0.03}
& \greencell \ci{0.03}
& \ci{0.16}
& \ci{0.09}
& \greencell \ci{0.03}
& \ci{0.04}
& \ci{0.03}
& \greencell \ci{0.03}
& \ci{0.11}
& \ci{2.80}
& \greencell \ci{0.02}
& \greencell \ci{0.03}
& \greencell \ci{0.02} \\
\cmidrule(lr){3-3}
\cmidrule(lr){4-16}

&
&
& \greencell \num{65.04}
& $\top$
& $\top$
& $\top$
& $\top$
& $\top$
& $\top$
& $\top$
& $\top$
& $\top$
& $\top$
& $\top$
& $\top$ \\
&
& \multirow{-2}{*}{TIF008}
& \greencell \ci{23.61}
& 
& 
& 
& 
& 
& 
& 
& 
& 
& 
& 
&  \\
\cmidrule(lr){3-3}
\cmidrule(lr){4-16}

&
&
& \num{28.49}
& \num{30.93}
& \num{25.45}
& \num{37.69}
& \num{33.09}
& \num{23.03}
& \num{18.79}
& \num{19.39}
& \greencell \num{3.29}
& \num{10.62}
& \num{14.31}
& \num{33.37}
& \num{33.77} \\
&
& \multirow{-2}{*}{TIF009}
& \ci{19.49}
& \ci{20.82}
& \ci{19.99}
& \ci{17.37}
& \ci{22.14}
& \ci{18.04}
& \ci{11.26}
& \ci{14.14}
& \greencell \ci{2.11}
& \ci{1.53}
& \ci{3.47}
& \ci{15.03}
& \ci{17.58} \\
\cmidrule(lr){3-3}
\cmidrule(lr){4-16}

&
&
& \greencell \num{1.26}
& \greencell \num{0.86}
& \greencell \num{1.33}
& \num{7.77}
& \num{2.41}
& \greencell \num{1.36}
& \greencell \num{0.89}
& \greencell \num{1.37}
& \greencell \num{7.30}
& \num{54.88}
& \num{2.43}
& \greencell \num{1.53}
& \greencell \num{1.15} \\
&
& \multirow{-2}{*}{TIF012}
& \greencell \ci{0.30}
& \greencell \ci{0.31}
& \greencell \ci{0.51}
& \ci{5.61}
& \ci{1.05}
& \greencell \ci{0.45}
& \greencell \ci{0.22}
& \greencell \ci{0.57}
& \greencell \ci{5.82}
& \ci{15.72}
& \ci{0.99}
& \greencell \ci{0.79}
& \greencell \ci{0.39} \\
\cmidrule(lr){3-3}
\cmidrule(lr){4-16}

&
&
& \num{4.06}
& \num{3.18}
& \num{1.80}
& \num{9.53}
& \num{3.93}
& \num{2.48}
& \greencell \num{1.32}
& \greencell \num{1.05}
& \num{5.68}
& \num{61.01}
& \greencell \num{1.29}
& \greencell \num{0.93}
& \greencell \num{0.87} \\
\multirow{-40}{*}{libtiff} & \multirow{-19}{*}{tiffcp} &  \multirow{-2}{*}{TIF014}
& \ci{1.99}
& \ci{1.49}
& \ci{0.60}
& \ci{7.82}
& \ci{2.29}
& \ci{1.06}
& \greencell \ci{0.43}
& \greencell \ci{0.33}
& \ci{2.66}
& \ci{15.90}
& \greencell \ci{0.61}
& \greencell \ci{0.44}
& \greencell \ci{0.39} \\
\bottomrule
\end{tabular}
\end{adjustbox}
\end{table*}

\begin{table*}\ContinuedFloat
\centering
\footnotesize
\caption{Magma bugs (cont.).}

\begin{adjustbox}{width=\linewidth}
\begin{tabular}{>{\itshape}l>{\ttfamily}ll *{13}{r}}
\toprule
& & & \multicolumn{13}{c}{Fuzzer} \\
\cmidrule(lr){4-16}
& & & \multicolumn{8}{c}{AFL++} & & & \multicolumn{3}{c}{\algoname} \\
\cmidrule(lr){4-11}
\cmidrule(lr){14-16}
\multirow{-3}{*}{\normalfont{Target}}
& \multirow{-3}{*}{\normalfont{Driver}}
& \multirow{-3}{*}{Bug}
& EXPLORE
& FAST
& COE
& QUAD
& LIN
& EXPLOIT
& MMOPT
& RARE
& \multirow{-2}{*}{\texttt{K-Sched}}
& \multirow{-2}{*}{Tortoise}
& \algoworare
& \algowrare
& \algosample \\
\midrule
&
&
& $\top$
& $\top$
& \num{67.43}
& $\top$
& $\top$
& $\top$
& \greencell \num{43.49}
& $\top$
& $\top$
& $\top$
& $\top$
& \num{65.80}
& \num{65.02} \\
&
& \multirow{-2}{*}{XML001}
& 
& 
& \ci{8.15}
& 
& 
& 
& \greencell \ci{14.41}
& 
& 
& 
& 
& \ci{8.42}
& \ci{13.91} \\
\cmidrule(lr){3-3}
\cmidrule(lr){4-16}

&
&
& $\top$
& $\top$
& $\top$
& $\top$
& \greencell \num{71.33}
& $\top$
& \greencell \num{65.73}
& \greencell \num{67.52}
& $\top$
& $\top$
& $\top$
& \greencell \num{68.72}
& \greencell \num{61.70} \\
&
& \multirow{-2}{*}{XML002}
& 
& 
& 
& 
& \greencell \ci{2.27}
& 
& \greencell \ci{21.28}
& \greencell \ci{15.20}
& 
& 
& 
& \greencell \ci{11.15}
& \greencell \ci{20.67} \\
\cmidrule(lr){3-3}
\cmidrule(lr){4-16}

&
&
& \num{5.49}
& \greencell \num{2.78}
& \greencell \num{2.59}
& \greencell \num{1.94}
& \greencell \num{2.63}
& \greencell \num{8.58}
& \greencell \num{9.29}
& \greencell \num{3.58}
& $\top$
& $\top$
& \num{4.93}
& \greencell \num{1.69}
& \greencell \num{2.84} \\
&
& \multirow{-2}{*}{XML003}
& \ci{2.49}
& \greencell \ci{2.09}
& \greencell \ci{0.92}
& \greencell \ci{1.16}
& \greencell \ci{0.80}
& \greencell \ci{5.46}
& \greencell \ci{12.41}
& \greencell \ci{1.82}
& 
& 
& \ci{2.74}
& \greencell \ci{0.83}
& \greencell \ci{1.21} \\
\cmidrule(lr){3-3}
\cmidrule(lr){4-16}

&
&
& \greencell \num{1.11}
& \greencell \num{1.52}
& \greencell \num{1.43}
& \num{2.45}
& \num{5.16}
& \num{4.83}
& \greencell \num{8.16}
& \greencell \num{1.82}
& $\top$
& $\top$
& \greencell \num{1.55}
& \greencell \num{1.64}
& \greencell \num{1.20} \\
&
& \multirow{-2}{*}{XML009}
& \greencell \ci{0.23}
& \greencell \ci{0.48}
& \greencell \ci{0.46}
& \ci{0.92}
& \ci{2.16}
& \ci{1.73}
& \greencell \ci{12.59}
& \greencell \ci{0.88}
& 
& 
& \greencell \ci{0.90}
& \greencell \ci{0.91}
& \greencell \ci{0.46} \\
\cmidrule(lr){3-3}
\cmidrule(lr){4-16}

&
&
& \num{69.16}
& \greencell \num{60.42}
& \num{70.18}
& $\top$
& \greencell \num{63.83}
& $\top$
& \greencell \num{48.18}
& $\top$
& $\top$
& $\top$
& $\top$
& $\top$
& \num{71.61} \\
&
& \multirow{-2}{*}{XML012}
& \ci{9.65}
& \greencell \ci{11.63}
& \ci{6.19}
& 
& \greencell \ci{12.93}
& 
& \greencell \ci{18.08}
& 
& 
& 
& 
& 
& \ci{1.33} \\
\cmidrule(lr){3-3}
\cmidrule(lr){4-16}

&
&
& \greencell \num{0.02}
& \greencell \num{0.02}
& \greencell \num{0.02}
& \greencell \num{0.04}
& \greencell \num{0.06}
& \greencell \num{0.02}
& \greencell \num{7.21}
& \greencell \num{0.03}
& \greencell \num{0.02}
& \greencell \num{0.03}
& \greencell \num{0.02}
& \greencell \num{0.02}
& \greencell \num{0.03} \\
& \multirow{-15}{*}{xml\_read\_memory\_fuzzer} &  \multirow{-2}{*}{XML017}
& \greencell \ci{0.02}
& \greencell \ci{0.02}
& \greencell \ci{0.02}
& \greencell \ci{0.06}
& \greencell \ci{0.04}
& \greencell \ci{0.02}
& \greencell \ci{16.00}
& \greencell \ci{0.02}
& \greencell \ci{0.02}
& \greencell \ci{0.03}
& \greencell \ci{0.02}
& \greencell \ci{0.01}
& \greencell \ci{0.02} \\
\cmidrule(lr){2-2}
\cmidrule(lr){3-3}
\cmidrule(lr){4-16}

&
&
& \greencell \num{58.72}
& \greencell \num{62.41}
& \greencell \num{63.36}
& \num{68.58}
& \greencell \num{60.06}
& $\top$
& \greencell \num{54.85}
& \greencell \num{65.02}
& $\top$
& $\top$
& \greencell \num{62.34}
& \greencell \num{52.02}
& \greencell \num{57.17} \\
&
& \multirow{-2}{*}{XML001}
& \greencell \ci{11.70}
& \greencell \ci{9.50}
& \greencell \ci{7.42}
& \ci{11.62}
& \greencell \ci{16.06}
& 
& \greencell \ci{11.93}
& \greencell \ci{10.46}
& 
& 
& \greencell \ci{8.09}
& \greencell \ci{11.68}
& \greencell \ci{8.30} \\
\cmidrule(lr){3-3}
\cmidrule(lr){4-16}

&
&
& \greencell \num{65.11}
& \greencell \num{71.07}
& \greencell \num{68.13}
& $\top$
& \greencell \num{66.00}
& $\top$
& $\top$
& \greencell \num{66.75}
& $\top$
& $\top$
& \greencell \num{69.56}
& \greencell \num{66.25}
& \greencell \num{65.02} \\
&
& \multirow{-2}{*}{XML002}
& \greencell \ci{14.82}
& \greencell \ci{3.16}
& \greencell \ci{13.14}
& 
& \greencell \ci{20.38}
& 
& 
& \greencell \ci{17.82}
& 
& 
& \greencell \ci{8.29}
& \greencell \ci{11.28}
& \greencell \ci{23.70} \\
\cmidrule(lr){3-3}
\cmidrule(lr){4-16}

&
&
& \num{1.47}
& \num{2.03}
& \num{2.01}
& \num{5.89}
& \num{6.37}
& \num{6.17}
& \num{2.30}
& \num{2.70}
& \num{66.68}
& $\top$
& \greencell \num{1.11}
& \greencell \num{0.93}
& \greencell \num{0.64} \\
&
& \multirow{-2}{*}{XML009}
& \ci{0.72}
& \ci{0.92}
& \ci{0.80}
& \ci{2.55}
& \ci{2.64}
& \ci{2.18}
& \ci{1.27}
& \ci{1.53}
& \ci{9.16}
& 
& \greencell \ci{0.40}
& \greencell \ci{0.46}
& \greencell \ci{0.21} \\
\cmidrule(lr){3-3}
\cmidrule(lr){4-16}

&
&
& $\top$
& $\top$
& \greencell \num{65.92}
& \greencell \num{65.67}
& \greencell \num{66.99}
& $\top$
& \greencell \num{65.99}
& $\top$
& $\top$
& $\top$
& $\top$
& $\top$
& \greencell \num{67.14} \\
&
& \multirow{-2}{*}{XML012}
& 
& 
& \greencell \ci{12.90}
& \greencell \ci{21.48}
& \greencell \ci{17.02}
& 
& \greencell \ci{20.39}
& 
& 
& 
& 
& 
& \greencell \ci{14.06} \\
\cmidrule(lr){3-3}
\cmidrule(lr){4-16}

&
&
& \greencell \num{0.03}
& \greencell \num{0.05}
& \greencell \num{0.04}
& \greencell \num{0.07}
& \greencell \num{0.06}
& \greencell \num{0.02}
& \num{0.03}
& \greencell \num{0.02}
& \greencell \num{0.01}
& \num{0.13}
& \greencell \num{0.04}
& \num{0.03}
& \greencell \num{0.03} \\
\multirow{-25}{*}{libxml2} & \multirow{-13}{*}{xmllint} &  \multirow{-2}{*}{XML017}
& \greencell \ci{0.02}
& \greencell \ci{0.05}
& \greencell \ci{0.03}
& \greencell \ci{0.07}
& \greencell \ci{0.04}
& \greencell \ci{0.02}
& \ci{0.02}
& \greencell \ci{0.02}
& \greencell \ci{0.02}
& \ci{0.09}
& \greencell \ci{0.03}
& \ci{0.02}
& \greencell \ci{0.02} \\
\cmidrule(lr){1-1}
\cmidrule(lr){2-2}
\cmidrule(lr){3-3}
\cmidrule(lr){4-16}
&
&
& $\top$
& $\top$
& $\top$
& $\top$
& $\top$
& $\top$
& $\top$
& $\top$
& $\top$
& $\top$
& \greencell \num{67.10}
& \greencell \num{69.76}
& \greencell \num{71.10} \\
&
& \multirow{-2}{*}{LUA002}
& 
& 
& 
& 
& 
& 
& 
& 
& 
& 
& \greencell \ci{6.58}
& \greencell \ci{7.61}
& \greencell \ci{3.04} \\
\cmidrule(lr){3-3}
\cmidrule(lr){4-16}

&
&
& \greencell \num{5.68}
& \greencell \num{8.15}
& \greencell \num{5.75}
& \num{14.95}
& \num{36.47}
& \num{35.36}
& \greencell \num{5.89}
& \num{10.19}
& \greencell \num{9.93}
& \greencell \num{7.21}
& \greencell \num{9.69}
& \greencell \num{6.24}
& \num{10.03} \\
\multirow{-4}{*}{lua} & \multirow{-4}{*}{lua} &  \multirow{-2}{*}{LUA004}
& \greencell \ci{2.17}
& \greencell \ci{2.27}
& \greencell \ci{2.87}
& \ci{5.97}
& \ci{20.63}
& \ci{9.31}
& \greencell \ci{3.57}
& \ci{4.25}
& \greencell \ci{4.11}
& \greencell \ci{17.28}
& \greencell \ci{2.90}
& \greencell \ci{2.08}
& \ci{2.58} \\
\cmidrule(lr){1-1}
\cmidrule(lr){2-2}
\cmidrule(lr){3-3}
\cmidrule(lr){4-16}
&
&
& \num{35.11}
& \num{25.39}
& \num{28.46}
& \num{44.71}
& \num{47.63}
& \greencell \num{8.58}
& \num{19.74}
& \num{38.69}
& \num{66.85}
& $\top$
& \greencell \num{5.72}
& \greencell \num{5.45}
& \greencell \num{6.53} \\
&
& \multirow{-2}{*}{SSL001}
& \ci{12.55}
& \ci{7.22}
& \ci{9.54}
& \ci{11.97}
& \ci{13.78}
& \greencell \ci{3.50}
& \ci{6.45}
& \ci{9.26}
& \ci{17.47}
& 
& \greencell \ci{2.27}
& \greencell \ci{2.84}
& \greencell \ci{3.68} \\
\cmidrule(lr){3-3}
\cmidrule(lr){4-16}

&
&
& \num{0.06}
& \num{0.06}
& \num{0.06}
& \num{0.06}
& \num{0.06}
& \num{0.06}
& \greencell \num{0.06}
& \num{0.06}
& \num{0.16}
& \num{0.26}
& \num{0.06}
& \num{0.07}
& \num{0.07} \\
& \multirow{-4}{*}{asn1} &  \multirow{-2}{*}{SSL003}
& \ci{0.07}
& \ci{0.06}
& \ci{0.06}
& \ci{0.06}
& \ci{0.06}
& \ci{0.05}
& \greencell \ci{0.05}
& \ci{0.05}
& \ci{0.00}
& \ci{0.00}
& \ci{0.04}
& \ci{0.08}
& \ci{0.07} \\
\cmidrule(lr){2-2}
\cmidrule(lr){3-3}
\cmidrule(lr){4-16}

&
&
& \greencell \num{0.08}
& \greencell \num{0.17}
& \greencell \num{0.07}
& \greencell \num{0.08}
& \greencell \num{0.08}
& \greencell \num{0.08}
& \greencell \num{0.07}
& \greencell \num{0.08}
& \num{0.17}
& \num{50.42}
& \num{0.09}
& \greencell \num{0.08}
& \greencell \num{0.09} \\
& \multirow{-2}{*}{client} &  \multirow{-2}{*}{SSL002}
& \greencell \ci{0.06}
& \greencell \ci{0.20}
& \greencell \ci{0.05}
& \greencell \ci{0.06}
& \greencell \ci{0.06}
& \greencell \ci{0.05}
& \greencell \ci{0.05}
& \greencell \ci{0.05}
& \ci{0.00}
& \ci{37.31}
& \ci{0.08}
& \greencell \ci{0.06}
& \greencell \ci{0.06} \\
\cmidrule(lr){2-2}
\cmidrule(lr){3-3}
\cmidrule(lr){4-16}

&
&
& \num{0.11}
& \greencell \num{0.11}
& \num{0.12}
& \num{0.16}
& \greencell \num{0.11}
& \num{0.12}
& \greencell \num{0.16}
& \greencell \num{0.11}
& \num{0.22}
& \num{0.35}
& \num{0.11}
& \num{0.11}
& \num{0.12} \\
&
& \multirow{-2}{*}{SSL002}
& \ci{0.08}
& \greencell \ci{0.08}
& \ci{0.08}
& \ci{0.09}
& \greencell \ci{0.08}
& \ci{0.08}
& \greencell \ci{0.09}
& \greencell \ci{0.08}
& \ci{0.00}
& \ci{0.00}
& \ci{0.08}
& \ci{0.08}
& \ci{0.09} \\
\cmidrule(lr){3-3}
\cmidrule(lr){4-16}

&
&
& $\top$
& $\top$
& $\top$
& $\top$
& $\top$
& $\top$
& $\top$
& $\top$
& \greencell \num{18.62}
& \greencell \num{16.42}
& \greencell \num{29.93}
& \num{37.10}
& \num{46.80} \\
& \multirow{-4}{*}{server} &  \multirow{-2}{*}{SSL020}
& 
& 
& 
& 
& 
& 
& 
& 
& \greencell \ci{4.02}
& \greencell \ci{3.27}
& \greencell \ci{16.92}
& \ci{14.20}
& \ci{16.06} \\
\cmidrule(lr){2-2}
\cmidrule(lr){3-3}
\cmidrule(lr){4-16}

&
&
& $\top$
& \num{71.49}
& \num{66.82}
& $\top$
& $\top$
& \num{64.89}
& $\top$
& \num{54.42}
& $\top$
& \greencell \num{27.31}
& $\top$
& $\top$
& $\top$ \\
\multirow{-15}{*}{openssl} & \multirow{-2}{*}{x509} &  \multirow{-2}{*}{SSL009}
& 
& \ci{1.74}
& \ci{17.60}
& 
& 
& \ci{12.55}
& 
& \ci{19.80}
& 
& \greencell \ci{17.28}
& 
& 
&  \\
\cmidrule(lr){1-1}
\cmidrule(lr){2-2}
\cmidrule(lr){3-3}
\cmidrule(lr){4-16}
&
&
& \num{57.62}
& \num{70.00}
& \num{49.60}
& \num{57.61}
& $\top$
& \num{48.32}
& \num{65.14}
& \num{51.48}
& \xmark
& \greencell \num{2.77}
& \greencell \num{5.61}
& \greencell \num{5.48}
& \greencell \num{2.88} \\
&
& \multirow{-2}{*}{PHP004}
& \ci{28.19}
& \ci{6.80}
& \ci{23.07}
& \ci{28.20}
& 
& \ci{16.34}
& \ci{23.29}
& \ci{27.52}
& 
& \greencell \ci{0.06}
& \greencell \ci{3.11}
& \greencell \ci{5.15}
& \greencell \ci{2.54} \\
\cmidrule(lr){3-3}
\cmidrule(lr){4-16}

&
&
& \num{56.61}
& \num{30.29}
& \num{49.65}
& \num{68.83}
& \num{61.50}
& \num{15.25}
& \num{27.63}
& \num{33.01}
& \xmark
& \num{3.51}
& \greencell \num{1.22}
& \greencell \num{0.64}
& \greencell \num{0.98} \\
&
& \multirow{-2}{*}{PHP009}
& \ci{17.72}
& \ci{17.40}
& \ci{24.01}
& \ci{8.99}
& \ci{14.04}
& \ci{7.36}
& \ci{19.74}
& \ci{20.78}
& 
& \ci{0.22}
& \greencell \ci{0.76}
& \greencell \ci{0.28}
& \greencell \ci{0.57} \\
\cmidrule(lr){3-3}
\cmidrule(lr){4-16}

&
&
& \num{2.55}
& \greencell \num{1.67}
& \num{3.16}
& \num{1.54}
& \num{3.80}
& \num{0.70}
& \num{1.42}
& \num{1.11}
& \xmark
& \num{2.23}
& \greencell \num{0.13}
& \greencell \num{0.21}
& \greencell \num{0.22} \\
\multirow{-7}{*}{php} & \multirow{-7}{*}{exif} &  \multirow{-2}{*}{PHP011}
& \ci{1.37}
& \greencell \ci{1.89}
& \ci{2.88}
& \ci{1.14}
& \ci{3.16}
& \ci{0.41}
& \ci{1.03}
& \ci{0.94}
& 
& \ci{0.03}
& \greencell \ci{0.06}
& \greencell \ci{0.07}
& \greencell \ci{0.09} \\
\cmidrule(lr){1-1}
\cmidrule(lr){2-2}
\cmidrule(lr){3-3}
\cmidrule(lr){4-16}
&
&
& \greencell \num{1.28}
& \greencell \num{2.28}
& \greencell \num{2.62}
& \num{9.57}
& \num{3.56}
& \num{3.70}
& \greencell \num{1.31}
& \greencell \num{1.21}
& \num{62.10}
& $\top$
& \num{2.83}
& \num{5.19}
& \num{2.77} \\
&
& \multirow{-2}{*}{SQL002}
& \greencell \ci{0.50}
& \greencell \ci{0.88}
& \greencell \ci{1.98}
& \ci{2.10}
& \ci{0.99}
& \ci{1.32}
& \greencell \ci{0.63}
& \greencell \ci{0.41}
& \ci{19.45}
& 
& \ci{1.26}
& \ci{1.63}
& \ci{1.09} \\
\cmidrule(lr){3-3}
\cmidrule(lr){4-16}

&
&
& $\top$
& \greencell \num{68.65}
& $\top$
& \greencell \num{68.44}
& \greencell \num{66.47}
& $\top$
& $\top$
& $\top$
& $\top$
& $\top$
& $\top$
& \greencell \num{69.81}
& \greencell \num{71.67} \\
&
& \multirow{-2}{*}{SQL003}
& 
& \greencell \ci{11.38}
& 
& \greencell \ci{12.09}
& \greencell \ci{18.78}
& 
& 
& 
& 
& 
& 
& \greencell \ci{7.45}
& \greencell \ci{1.13} \\
\cmidrule(lr){3-3}
\cmidrule(lr){4-16}

&
&
& $\top$
& $\top$
& $\top$
& $\top$
& \greencell \num{68.12}
& $\top$
& \greencell \num{70.78}
& $\top$
& $\top$
& $\top$
& \greencell \num{66.87}
& $\top$
& $\top$ \\
&
& \multirow{-2}{*}{SQL010}
& 
& 
& 
& 
& \greencell \ci{13.19}
& 
& \greencell \ci{4.15}
& 
& 
& 
& \greencell \ci{17.42}
& 
&  \\
\cmidrule(lr){3-3}
\cmidrule(lr){4-16}

&
&
& \greencell \num{48.45}
& \greencell \num{56.60}
& \greencell \num{63.50}
& $\top$
& \greencell \num{54.90}
& $\top$
& \greencell \num{61.02}
& \greencell \num{60.32}
& $\top$
& $\top$
& \num{67.25}
& \greencell \num{63.18}
& \greencell \num{54.53} \\
&
& \multirow{-2}{*}{SQL012}
& \greencell \ci{14.52}
& \greencell \ci{10.60}
& \greencell \ci{13.57}
& 
& \greencell \ci{20.57}
& 
& \greencell \ci{9.13}
& \greencell \ci{13.18}
& 
& 
& \ci{9.35}
& \greencell \ci{15.02}
& \greencell \ci{23.44} \\
\cmidrule(lr){3-3}
\cmidrule(lr){4-16}

&
&
& $\top$
& \greencell \num{67.15}
& \greencell \num{69.68}
& $\top$
& \greencell \num{69.31}
& $\top$
& $\top$
& $\top$
& $\top$
& $\top$
& \greencell \num{71.16}
& \greencell \num{67.38}
& \greencell \num{62.88} \\
&
& \multirow{-2}{*}{SQL013}
& 
& \greencell \ci{8.35}
& \greencell \ci{7.89}
& 
& \greencell \ci{7.06}
& 
& 
& 
& 
& 
& \greencell \ci{2.86}
& \greencell \ci{9.07}
& \greencell \ci{13.30} \\
\cmidrule(lr){3-3}
\cmidrule(lr){4-16}

&
&
& \greencell \num{8.63}
& \greencell \num{8.64}
& \num{17.78}
& \num{44.40}
& \greencell \num{18.42}
& \greencell \num{17.90}
& \greencell \num{19.91}
& \num{30.75}
& $\top$
& $\top$
& \greencell \num{13.94}
& \num{29.72}
& \greencell \num{15.60} \\
&
& \multirow{-2}{*}{SQL014}
& \greencell \ci{4.36}
& \greencell \ci{2.56}
& \ci{6.82}
& \ci{13.27}
& \greencell \ci{9.06}
& \greencell \ci{9.91}
& \greencell \ci{11.24}
& \ci{10.16}
& 
& 
& \greencell \ci{4.39}
& \ci{10.17}
& \greencell \ci{7.58} \\
\cmidrule(lr){3-3}
\cmidrule(lr){4-16}

&
&
& \greencell \num{70.67}
& \greencell \num{64.43}
& \greencell \num{67.36}
& $\top$
& \greencell \num{57.17}
& $\top$
& \greencell \num{66.12}
& \greencell \num{64.72}
& $\top$
& $\top$
& $\top$
& \greencell \num{69.17}
& \greencell \num{66.67} \\
&
& \multirow{-2}{*}{SQL015}
& \greencell \ci{4.50}
& \greencell \ci{14.97}
& \greencell \ci{15.75}
& 
& \greencell \ci{22.20}
& 
& \greencell \ci{12.17}
& \greencell \ci{14.34}
& 
& 
& 
& \greencell \ci{9.61}
& \greencell \ci{14.13} \\
\cmidrule(lr){3-3}
\cmidrule(lr){4-16}

&
&
& \greencell \num{4.60}
& \greencell \num{3.98}
& \greencell \num{8.58}
& \num{19.84}
& \greencell \num{4.72}
& \num{12.69}
& \greencell \num{3.40}
& \greencell \num{3.90}
& $\top$
& $\top$
& \greencell \num{5.64}
& \greencell \num{5.41}
& \greencell \num{6.21} \\
&
& \multirow{-2}{*}{SQL018}
& \greencell \ci{1.56}
& \greencell \ci{1.64}
& \greencell \ci{4.84}
& \ci{10.26}
& \greencell \ci{1.11}
& \ci{4.12}
& \greencell \ci{1.66}
& \greencell \ci{1.99}
& 
& 
& \greencell \ci{2.30}
& \greencell \ci{1.50}
& \greencell \ci{1.59} \\
\cmidrule(lr){3-3}
\cmidrule(lr){4-16}

&
&
& \greencell \num{42.36}
& \greencell \num{46.39}
& \num{60.29}
& \num{69.81}
& \greencell \num{40.07}
& \greencell \num{55.64}
& \greencell \num{55.97}
& \num{67.64}
& $\top$
& $\top$
& \num{61.24}
& \greencell \num{59.17}
& \num{64.01} \\
\multirow{-22}{*}{sqlite3} & \multirow{-22}{*}{sqlite3\_fuzz} &  \multirow{-2}{*}{SQL020}
& \greencell \ci{12.23}
& \greencell \ci{14.82}
& \ci{15.71}
& \ci{7.45}
& \greencell \ci{14.72}
& \greencell \ci{21.97}
& \greencell \ci{18.59}
& \ci{7.93}
& 
& 
& \ci{21.57}
& \greencell \ci{15.05}
& \ci{11.64} \\
\bottomrule
\end{tabular}
\end{adjustbox}
\end{table*}

\begin{table*}\ContinuedFloat
\centering
\footnotesize
\caption{Magma bugs (cont.).}

\begin{adjustbox}{width=\linewidth}
\begin{tabular}{>{\itshape}l>{\ttfamily}ll *{13}{r}}
\toprule
& & & \multicolumn{13}{c}{Fuzzer} \\
\cmidrule(lr){4-16}
& & & \multicolumn{8}{c}{AFL++} & & & \multicolumn{3}{c}{\algoname} \\
\cmidrule(lr){4-11}
\cmidrule(lr){14-16}
\multirow{-3}{*}{\normalfont{Target}}
& \multirow{-3}{*}{\normalfont{Driver}}
& \multirow{-3}{*}{Bug}
& EXPLORE
& FAST
& COE
& QUAD
& LIN
& EXPLOIT
& MMOPT
& RARE
& \multirow{-2}{*}{\texttt{K-Sched}}
& \multirow{-2}{*}{Tortoise}
& \algoworare
& \algowrare
& \algosample \\
\midrule
&
&
& $\top$
& \greencell \num{65.08}
& $\top$
& $\top$
& $\top$
& $\top$
& $\top$
& $\top$
& \xmark
& $\top$
& $\top$
& $\top$
& $\top$ \\
&
& \multirow{-2}{*}{PDF001}
& 
& \greencell \ci{23.48}
& 
& 
& 
& 
& 
& 
& 
& 
& 
& 
&  \\
\cmidrule(lr){3-3}
\cmidrule(lr){4-16}

&
&
& \num{1.15}
& \num{1.82}
& \num{1.89}
& \num{5.25}
& \num{5.63}
& \num{2.07}
& \num{1.96}
& \num{1.61}
& \xmark
& \greencell \num{0.10}
& \num{0.99}
& \num{1.23}
& \num{1.24} \\
&
& \multirow{-2}{*}{PDF010}
& \ci{0.53}
& \ci{0.50}
& \ci{1.34}
& \ci{2.81}
& \ci{2.40}
& \ci{2.03}
& \ci{1.22}
& \ci{0.56}
& 
& \greencell \ci{0.10}
& \ci{0.47}
& \ci{0.52}
& \ci{0.69} \\
\cmidrule(lr){3-3}
\cmidrule(lr){4-16}

&
&
& \greencell \num{65.59}
& $\top$
& \greencell \num{66.53}
& \greencell \num{60.79}
& $\top$
& \greencell \num{65.88}
& $\top$
& $\top$
& \xmark
& $\top$
& \greencell \num{67.01}
& \greencell \num{65.70}
& \greencell \num{55.79} \\
&
& \multirow{-2}{*}{PDF011}
& \greencell \ci{21.76}
& 
& \greencell \ci{18.57}
& \greencell \ci{21.97}
& 
& \greencell \ci{20.79}
& 
& 
& 
& 
& \greencell \ci{12.96}
& \greencell \ci{21.39}
& \greencell \ci{21.95} \\
\cmidrule(lr){3-3}
\cmidrule(lr){4-16}

&
&
& \greencell \num{0.04}
& \greencell \num{0.05}
& \greencell \num{0.06}
& \greencell \num{0.07}
& \greencell \num{0.03}
& \greencell \num{0.04}
& \greencell \num{0.04}
& \greencell \num{0.07}
& \xmark
& \num{0.25}
& \greencell \num{0.04}
& \greencell \num{0.04}
& \greencell \num{0.05} \\
&
& \multirow{-2}{*}{PDF016}
& \greencell \ci{0.02}
& \greencell \ci{0.03}
& \greencell \ci{0.04}
& \greencell \ci{0.09}
& \greencell \ci{0.02}
& \greencell \ci{0.02}
& \greencell \ci{0.02}
& \greencell \ci{0.04}
& 
& \ci{0.00}
& \greencell \ci{0.02}
& \greencell \ci{0.02}
& \greencell \ci{0.03} \\
\cmidrule(lr){3-3}
\cmidrule(lr){4-16}

&
&
& \num{37.84}
& \num{40.38}
& \num{38.25}
& $\top$
& $\top$
& \num{33.83}
& \num{29.91}
& \greencell \num{20.92}
& \xmark
& $\top$
& \greencell \num{12.75}
& \greencell \num{9.40}
& \greencell \num{10.99} \\
&
& \multirow{-2}{*}{PDF018}
& \ci{22.46}
& \ci{20.71}
& \ci{19.84}
& 
& 
& \ci{13.80}
& \ci{16.76}
& \greencell \ci{12.37}
& 
& 
& \greencell \ci{6.18}
& \greencell \ci{4.68}
& \greencell \ci{5.44} \\
\cmidrule(lr){3-3}
\cmidrule(lr){4-16}

&
&
& $\top$
& $\top$
& $\top$
& $\top$
& \greencell \num{69.39}
& \greencell \num{62.62}
& $\top$
& $\top$
& \xmark
& $\top$
& $\top$
& $\top$
& $\top$ \\
&
& \multirow{-2}{*}{PDF019}
& 
& 
& 
& 
& \greencell \ci{8.85}
& \greencell \ci{21.37}
& 
& 
& 
& 
& 
& 
&  \\
\cmidrule(lr){3-3}
\cmidrule(lr){4-16}

&
&
& \greencell \num{52.56}
& $\top$
& $\top$
& \greencell \num{62.32}
& \greencell \num{55.67}
& $\top$
& \greencell \num{60.34}
& \greencell \num{65.11}
& \xmark
& $\top$
& \greencell \num{70.08}
& \greencell \num{68.57}
& \greencell \num{68.76} \\
& \multirow{-17}{*}{pdf\_fuzzer} &  \multirow{-2}{*}{PDF021}
& \greencell \ci{19.39}
& 
& 
& \greencell \ci{13.10}
& \greencell \ci{18.47}
& 
& \greencell \ci{23.04}
& \greencell \ci{23.38}
& 
& 
& \greencell \ci{6.50}
& \greencell \ci{11.63}
& \greencell \ci{10.99} \\
\cmidrule(lr){2-2}
\cmidrule(lr){3-3}
\cmidrule(lr){4-16}

&
&
& $\top$
& $\top$
& \greencell \num{65.84}
& $\top$
& $\top$
& $\top$
& $\top$
& $\top$
& \xmark
& $\top$
& $\top$
& \greencell \num{65.56}
& \greencell \num{65.57} \\
&
& \multirow{-2}{*}{PDF002}
& 
& 
& \greencell \ci{20.92}
& 
& 
& 
& 
& 
& 
& 
& 
& \greencell \ci{21.87}
& \greencell \ci{21.84} \\
\cmidrule(lr){3-3}
\cmidrule(lr){4-16}

&
&
& \greencell \num{10.42}
& \num{11.24}
& \greencell \num{7.80}
& \num{13.40}
& \greencell \num{9.72}
& \num{32.29}
& \num{31.47}
& \num{31.91}
& \xmark
& $\top$
& \num{23.56}
& \greencell \num{5.98}
& \greencell \num{9.75} \\
&
& \multirow{-2}{*}{PDF003}
& \greencell \ci{5.69}
& \ci{4.53}
& \greencell \ci{2.47}
& \ci{5.75}
& \greencell \ci{3.65}
& \ci{18.22}
& \ci{16.99}
& \ci{18.48}
& 
& 
& \ci{11.40}
& \greencell \ci{2.64}
& \greencell \ci{4.05} \\
\cmidrule(lr){3-3}
\cmidrule(lr){4-16}

&
&
& \num{67.30}
& \greencell \num{47.78}
& \greencell \num{50.65}
& \num{64.93}
& $\top$
& \num{70.10}
& \greencell \num{59.23}
& \greencell \num{56.30}
& \xmark
& \greencell \num{48.95}
& \greencell \num{55.77}
& \num{65.02}
& \greencell \num{35.84} \\
&
& \multirow{-2}{*}{PDF011}
& \ci{15.96}
& \greencell \ci{23.75}
& \greencell \ci{21.96}
& \ci{24.01}
& 
& \ci{6.46}
& \greencell \ci{18.00}
& \greencell \ci{22.15}
& 
& \greencell \ci{13.93}
& \greencell \ci{22.48}
& \ci{15.35}
& \greencell \ci{17.81} \\
\cmidrule(lr){3-3}
\cmidrule(lr){4-16}

&
&
& \num{0.03}
& \greencell \num{0.01}
& \num{0.03}
& \greencell \num{0.02}
& \greencell \num{0.03}
& \num{0.02}
& \num{0.03}
& \greencell \num{0.02}
& \xmark
& \num{0.09}
& \num{0.04}
& \num{0.03}
& \num{0.02} \\
&
& \multirow{-2}{*}{PDF016}
& \ci{0.02}
& \greencell \ci{0.01}
& \ci{0.02}
& \greencell \ci{0.01}
& \greencell \ci{0.02}
& \ci{0.01}
& \ci{0.02}
& \greencell \ci{0.01}
& 
& \ci{0.06}
& \ci{0.03}
& \ci{0.02}
& \ci{0.02} \\
\cmidrule(lr){3-3}
\cmidrule(lr){4-16}

&
&
& \num{15.29}
& \greencell \num{10.03}
& \num{12.76}
& \num{62.63}
& \num{68.55}
& \num{17.24}
& \greencell \num{5.49}
& \greencell \num{7.89}
& \xmark
& $\top$
& \greencell \num{4.86}
& \greencell \num{5.23}
& \greencell \num{3.85} \\
&
& \multirow{-2}{*}{PDF018}
& \ci{9.90}
& \greencell \ci{5.12}
& \ci{3.98}
& \ci{14.63}
& \ci{9.41}
& \ci{8.60}
& \greencell \ci{3.25}
& \greencell \ci{8.87}
& 
& 
& \greencell \ci{1.36}
& \greencell \ci{1.38}
& \greencell \ci{1.57} \\
\cmidrule(lr){3-3}
\cmidrule(lr){4-16}

&
&
& \greencell \num{59.02}
& \greencell \num{46.57}
& \greencell \num{59.70}
& \greencell \num{64.94}
& $\top$
& \greencell \num{65.11}
& \greencell \num{65.89}
& \greencell \num{67.23}
& \xmark
& $\top$
& \greencell \num{59.00}
& \greencell \num{59.37}
& $\top$ \\
&
& \multirow{-2}{*}{PDF019}
& \greencell \ci{25.54}
& \greencell \ci{21.60}
& \greencell \ci{24.13}
& \greencell \ci{23.96}
& 
& \greencell \ci{23.39}
& \greencell \ci{9.77}
& \greencell \ci{10.93}
& 
& 
& \greencell \ci{25.48}
& \greencell \ci{24.76}
&  \\
\cmidrule(lr){3-3}
\cmidrule(lr){4-16}

&
&
& \greencell \num{68.11}
& \greencell \num{56.31}
& \greencell \num{57.63}
& \greencell \num{53.10}
& \greencell \num{64.80}
& \greencell \num{60.48}
& \greencell \num{60.53}
& $\top$
& \xmark
& $\top$
& $\top$
& $\top$
& $\top$ \\
& \multirow{-17}{*}{pdfimages} &  \multirow{-2}{*}{PDF021}
& \greencell \ci{7.83}
& \greencell \ci{22.81}
& \greencell \ci{20.14}
& \greencell \ci{19.80}
& \greencell \ci{11.22}
& \greencell \ci{17.74}
& \greencell \ci{16.80}
& 
& 
& 
& 
& 
&  \\
\cmidrule(lr){2-2}
\cmidrule(lr){3-3}
\cmidrule(lr){4-16}

&
&
& $\top$
& \greencell \num{69.18}
& $\top$
& $\top$
& $\top$
& \greencell \num{66.84}
& \greencell \num{70.95}
& $\top$
& \xmark
& $\top$
& $\top$
& $\top$
& $\top$ \\
&
& \multirow{-2}{*}{PDF002}
& 
& \greencell \ci{9.57}
& 
& 
& 
& \greencell \ci{17.53}
& \greencell \ci{3.55}
& 
& 
& 
& 
& 
&  \\
\cmidrule(lr){3-3}
\cmidrule(lr){4-16}

&
&
& $\top$
& $\top$
& \greencell \num{66.15}
& $\top$
& $\top$
& $\top$
& $\top$
& $\top$
& \xmark
& $\top$
& $\top$
& $\top$
& $\top$ \\
&
& \multirow{-2}{*}{PDF004}
& 
& 
& \greencell \ci{12.04}
& 
& 
& 
& 
& 
& 
& 
& 
& 
&  \\
\cmidrule(lr){3-3}
\cmidrule(lr){4-16}

&
&
& \greencell \num{37.74}
& \greencell \num{47.02}
& \greencell \num{39.42}
& $\top$
& \num{67.36}
& \num{62.16}
& \greencell \num{43.73}
& \greencell \num{57.99}
& \xmark
& $\top$
& \num{65.15}
& \num{68.07}
& \num{69.96} \\
&
& \multirow{-2}{*}{PDF006}
& \greencell \ci{16.73}
& \greencell \ci{17.98}
& \greencell \ci{19.31}
& 
& \ci{15.77}
& \ci{19.31}
& \greencell \ci{15.04}
& \greencell \ci{27.47}
& 
& 
& \ci{13.44}
& \ci{7.54}
& \ci{6.93} \\
\cmidrule(lr){3-3}
\cmidrule(lr){4-16}

&
&
& \num{3.21}
& \num{2.98}
& \num{2.51}
& \num{3.79}
& \num{4.14}
& \num{2.79}
& \num{3.01}
& \num{2.08}
& \xmark
& \greencell \num{0.11}
& \num{0.87}
& \num{0.81}
& \num{1.15} \\
&
& \multirow{-2}{*}{PDF010}
& \ci{1.70}
& \ci{1.53}
& \ci{0.90}
& \ci{1.56}
& \ci{2.63}
& \ci{1.96}
& \ci{1.40}
& \ci{0.82}
& 
& \greencell \ci{0.08}
& \ci{0.82}
& \ci{0.41}
& \ci{0.48} \\
\cmidrule(lr){3-3}
\cmidrule(lr){4-16}

&
&
& \greencell \num{61.79}
& $\top$
& \greencell \num{51.66}
& \greencell \num{68.18}
& \greencell \num{54.37}
& \greencell \num{64.07}
& \greencell \num{59.46}
& \greencell \num{62.30}
& \xmark
& $\top$
& \greencell \num{66.46}
& \greencell \num{61.80}
& \greencell \num{55.98} \\
&
& \multirow{-2}{*}{PDF011}
& \greencell \ci{20.01}
& 
& \greencell \ci{27.29}
& \greencell \ci{12.97}
& \greencell \ci{24.48}
& \greencell \ci{16.67}
& \greencell \ci{19.30}
& \greencell \ci{19.04}
& 
& 
& \greencell \ci{18.79}
& \greencell \ci{20.31}
& \greencell \ci{22.97} \\
\cmidrule(lr){3-3}
\cmidrule(lr){4-16}

&
&
& \num{0.07}
& \greencell \num{0.03}
& \num{0.03}
& \greencell \num{0.02}
& \num{0.03}
& \greencell \num{0.04}
& \greencell \num{0.03}
& \greencell \num{0.03}
& \xmark
& \num{0.19}
& \num{0.04}
& \greencell \num{0.07}
& \num{0.04} \\
&
& \multirow{-2}{*}{PDF016}
& \ci{0.04}
& \greencell \ci{0.02}
& \ci{0.02}
& \greencell \ci{0.02}
& \ci{0.02}
& \greencell \ci{0.02}
& \greencell \ci{0.02}
& \greencell \ci{0.02}
& 
& \ci{0.00}
& \ci{0.04}
& \greencell \ci{0.07}
& \ci{0.03} \\
\cmidrule(lr){3-3}
\cmidrule(lr){4-16}

&
&
& \num{29.16}
& \greencell \num{22.78}
& \num{21.64}
& \num{65.46}
& \num{65.66}
& \num{61.72}
& \num{24.27}
& \num{22.05}
& \xmark
& $\top$
& \greencell \num{8.02}
& \greencell \num{7.30}
& \greencell \num{8.73} \\
&
& \multirow{-2}{*}{PDF018}
& \ci{14.25}
& \greencell \ci{16.31}
& \ci{6.97}
& \ci{22.20}
& \ci{21.51}
& \ci{17.43}
& \ci{12.33}
& \ci{8.44}
& 
& 
& \greencell \ci{5.23}
& \greencell \ci{2.37}
& \greencell \ci{2.30} \\
\cmidrule(lr){3-3}
\cmidrule(lr){4-16}

&
&
& \greencell \num{66.98}
& $\top$
& \greencell \num{69.24}
& $\top$
& \greencell \num{65.84}
& $\top$
& $\top$
& \greencell \num{64.85}
& \xmark
& $\top$
& \greencell \num{66.97}
& \greencell \num{69.87}
& $\top$ \\
&
& \multirow{-2}{*}{PDF019}
& \greencell \ci{17.05}
& 
& \greencell \ci{9.37}
& 
& \greencell \ci{12.95}
& 
& 
& \greencell \ci{24.28}
& 
& 
& \greencell \ci{17.06}
& \greencell \ci{7.24}
&  \\
\cmidrule(lr){3-3}
\cmidrule(lr){4-16}

&
&
& \greencell \num{49.11}
& \greencell \num{48.91}
& \greencell \num{56.02}
& \greencell \num{47.02}
& \greencell \num{64.56}
& \greencell \num{54.78}
& \greencell \num{42.11}
& \num{66.85}
& \xmark
& $\top$
& \greencell \num{52.93}
& \greencell \num{63.05}
& \greencell \num{56.91} \\
\multirow{-45}{*}{poppler} & \multirow{-22}{*}{pdftoppm} &  \multirow{-2}{*}{PDF021}
& \greencell \ci{22.90}
& \greencell \ci{12.70}
& \greencell \ci{16.93}
& \greencell \ci{16.10}
& \greencell \ci{11.22}
& \greencell \ci{20.24}
& \greencell \ci{18.53}
& \ci{11.22}
& 
& 
& \greencell \ci{18.80}
& \greencell \ci{13.16}
& \greencell \ci{21.43} \\
\bottomrule
\end{tabular}
\end{adjustbox}
\end{table*}

\end{document}